\documentclass[aps,prb,twocolumn,showpacs,twoside]{revtex4}
\usepackage[utf8]{inputenc}
\usepackage{theorem,graphicx,amssymb,amsmath,amsbsy,ifpdf,subfigure,dsfont,pxfonts,ifthen}
\newboolean{submitToPRB} \setboolean{submitToPRB}{false}
\newtheorem{thma}{Theorem}
\newcommand{\mycaption}[1]{\ifthenelse{\boolean{submitToPRB}}{\caption{(Color online) #1}}{\caption{#1}}}
\usepackage[unicode]{hyperref}
\let\labOrigRightarrow=\Rightarrow   \RequirePackage{marvosym}   \let\Rightarrow=\labOrigRightarrow
\ifpdf \usepackage{color}
\hypersetup{pdfstartview=FitH,pdfauthor={Oleg CHALAEV and Daniel Loss <shalaev.oleg@unibas.ch>, Department of Physics and Astronomy, University of Basel, Klingelbergstrasse 82, Basel, CH-4056, Switzerland}}%
\else \usepackage[dvips]{color} \fi
\newcommand{\pf}{p_F}
\newcommand{\bes}{\begin{equation*}}\newcommand{\ees}{\end{equation*}}
\newcommand{\bea}{\begin{eqnarray}} \newcommand{\eea}{\end{eqnarray}}
\newcommand{\beas}{\begin{eqnarray*}} \newcommand{\eeas}{\end{eqnarray*}}
\newcommand{\GR}{G_R}\newcommand{\GA}{G_A}\newcommand{\GRA}{G_{\mathrm{R/A}}}

\newcommand{\GRT}{G_R^T}\newcommand{\GAT}{G_A^T}

\newcommand{\GRE}{G^E_R}
\newcommand{\GAEw}{G^{E-\omega}_\mathrm A}
\newcommand{\hGR}{{\hat G}_R}\newcommand{\hGA}{{\hat G}_A} \newcommand{\hGK}{{\hat G}_K} 
\newcommand{\hGRE}{{\hat G}^E_R} \newcommand{\hGAE}{{\hat G}^E_A}
\newcommand{\hGREw}{{\hat G}^{E-\omega}_R} \newcommand{\hGAEw}{{\hat G}^{E-\omega}_A}
\newcommand{\Sp}{\mathop{\mathrm{Tr}}}\newcommand{\ud}{\mathrm d}
\newcommand{\arccot}{\mathrm{arcctg}}
\newcommand{\gr}{g_r}\newcommand{\ga}{g_a}
\newcommand{\gre}{g_r^E}\newcommand{\gae}{g_a^E}\newcommand{\grae}{g_{\mathrm{r/a}}^E}
\newcommand{\hgre}{{\hat g}_r^E}\newcommand{\hgae}{{\hat g}_a^E}

\newcommand{\vf}{v_\mathrm F}
\newcommand{\ipd}{\int\frac{\ud^2p}{(2\pi\hbar)^2}}\newcommand{\ippd}{\int\frac{\ud^2p'}{(2\pi\hbar)^2}}
\newcommand{\ipds}{\int\ud^2p/(2\pi\hbar)^2}
\newcommand{\ikd}{\int\frac{\ud^2k}{(2\pi\hbar)^2}}\newcommand{\ikds}{\int\ud^2k/(2\pi\hbar)^2}
\newcommand{\iqd}{\int\frac{\ud^2q}{(2\pi\hbar)^2}}\newcommand{\iqds}{\int\ud^2q/(2\pi\hbar)^2}

\newcommand{\ixi}{\int_{-\infty}^\infty\ud\xi} \newcommand{\iE}{\int_{-\infty}^\infty\frac{\ud E}{2\pi}}
\newcommand{\iEs}{\int_{-\infty}^\infty\ud E/(2\pi)}
\newcommand{\insl}[1]{\ifpdf \href{#1}{#1}\else {\tt #1}\fi}

\newcommand{\inslna}[2]{\ifpdf \href{#1}{#2}\else #2\fi}
\newcommand{\condmat}[1]{\ifpdf \href{http://de.arxiv.org/abs/cond-mat/?#1}{\tt cond-mat/#1}\else {\tt cond-mat/#1}\fi}

\newcommand{\elib}[1]{\ifpdf \href{file://#1}{\includegraphics[height=.5cm]{dvd}}\fi}
\newcommand{\mybf}[1]{{\bf #1}}\renewcommand{\vec}{\mybf}

\newcommand{\inslcm}[2]{\ifpdf \href{#1}{#2}\else #1\fi}

\begin{document}
\title{Spin orbit-induced anisotropic conductivity of a disordered 2DEG.}

\author{Oleg Chalaev$^{1,2}$ and Daniel Loss$^{1}$} 
\affiliation{$^{1}$Department of Physics, University of Basel, Klingelbergstrasse 82, Basel, CH-4056, Switzerland}
\affiliation{$^{2}$Department of Physics, University of Missouri-Columbia, Columbia, Missouri 65211, USA}
%\date{\today}
\date{February 18, 2009}
\ifpdf
\hypersetup{pdfstartview=FitH,% <-- when opened in Acrobat Reader, starts in "page-screen-wide" mode
pdftitle={Anisotropic conductivity of a diffusive matter in the presence of Rashba and Dresselhaus spin-orbit coupling},%
pdfsubject={How Rashba and Dresselhaus spin-orbit interaction affects charge conductivity. Calculation using disorder-averaging diagrammatic technique},%
pdfauthor={Oleg CHALAEV <chalaev@gmail.com> and Daniel Loss,
Department of Physics and Astronomy, University of Basel, Klingelbergstrasse 82, Basel, CH-4056, Switzerland},%
pdfkeywords={mesoscopics, disorder averaging, diffuson, diffusion approximation, Rashba,Dresselhaus, spin-orbit}}%
\fi
\begin{abstract}
We present a semi-automated computer-assisted method to generate and calculate diagrams in the disorder averaging approach to disordered 2D conductors with intrinsic spin-orbit
interaction (SOI). As an application, we calculate the effect of the SOI on the  charge conductivity  
for disordered 2D systems and rings in the presence of Rashba and Dresselhaus SOI.
In an infinite-size 2D system, anisotropic corrections to the conductivity tensor arise due to phase-coherence and the interplay of Rashba and Dresselhaus SOI.
The effect is more pronounced in the quasi-onedimensional case, where the conductivity becomes anisotropic already in the presence of only one type of SOI.
The anisotropy further increases if the time-reversal symmetry of the Hamiltonian is broken.
\end{abstract}
 % <-- put spaces in this list
\maketitle
\section{Introduction}
Spin-orbit interaction (SOI) is a promising tool for manipulating spin degrees of freedom via electric field;
because of that, it plays an important role in various novel microelectronic devices.\cite{spintronics}
In 2D semiconductor microstructures, Rashba\cite{RashbaSOI} and Dresselhaus\cite{Dresselhaus} types of SOI are the most important ones.
In phase-coherent diffusive systems, their dominant effect on the transport properties is the so-called anti-localization\cite{Zumbuhl,Skvortsov} -- isotropic
SOI-induced correction to the conductivity tensor, leading to the sign change of the phase-coherent correction to the conductivity.
Rashba and Dresselhaus SOI terms equally and independently contribute to the weak anti-localization correction.
The anisotropic contribution to the conductivity tensor is a more subtle effect (arising in the next order in the weak-disorder expansion).
In an infinite-size 2D conductor it comes from the interference between Rashba and Dresselhaus SOI.\cite{anisotrCond}

A pure Dresselhaus SOI affects 2D electron systems in the same way as pure Rashba SOI with the same amplitude.
However, it is known that in a system with mixed (Rashba-Dresselhaus) type of SOI their action is not independent.
This becomes most pronounced in the special case when the amplitude of Rashba SOI is equal to the Dresselhaus one.\cite{schliemann03:146801,anisotrCond}
In order to highlight the interference between the Rashba and Dresselhaus SOI, it is convenient to consider effects,
which arise entirely due to this interference.
An example is the anisotropic contribution to the conductivity of a diffusive (unconfined) 2D electron gas, which is zero in case when only one type of SOI
is present in the system.\cite{anisotrCond}

Both the (isotropic) antilocalization correction and the SOI-induced anisotropic correction are phase coherent effects; hence
it is not surprising that \emph{in a fully phase coherent system} both of them depend singularly on SOI amplitudes.
In the limit of small SOI the weak localization correction diverges both in 2D and quasi-1D cases, while the behavior of the anisotropic correction depends on system's geometry:
in an infinite 2D-system with time-reversal symmetry it remains finite also when SOI is infinitesimal, and in a quasi-1D case it diverges with the vanishing~SOI.
Moreover, while in an infinite 2D disordered slab, two different
types of SOI (Rashba and Dresselhaus) are required in order to make the anisotropic component of the conductivity tensor non-zero\cite{anisotrCond}, in
a quasi-1D geometry (see Sec.~\ref{sec:quasi1D}) the SOI-induced anisotropy of the conductivity tensor arises also for the case when the energy spectrum is
isotropic (i.e., when only one type of SOI is present), despite the fact that all dimensions of a quasi-1D sample are much larger than the mean-free path $l$
of an electron.
Thus, in the phase-coherent regime the macroscopic shape anisotropy of the sample results in the (microscopic\cite{note:noDisp}) anisotropy of the
conductivity tensor. The SOI is still required, but the energy spectrum does not need to be anisotropic.

According to the theorem by Vollhardt and Wölfle (proved in\cite{woelfe} for the spinless case),  diffuson propagator poles do not contribute
  to the conductivity if the  system is invariant under time reversal.
We extend the validity of this theorem to the spinful case in Appendix~\ref{app:VW}.
From this theorem one may expect the appearance of uncompensated diffuson divergences in systems with broken time reversal symmetry, which would then
result in enhancement of the SOI-correction to the conductivity tensor.
We indeed observe such an enhancement in the example of a ring pierced by a magnetic flux, where the time-reversal symmetry is broken due to presence of the vector
potential (see Sec.~\ref{sec:brokenTR}).

We perform our calculations using the disorder averaging diagrammatic techniques.\cite{AGD,montambauxBook,Bergmann,DHIKSZ}
In problems with spin, a summation over spin indices produces huge expressions which cannot be  handled
manually anymore in a reasonable time. We have overcome this problem by
developing a symbolic-calculation  program \cite{theProgram} that (i) generates diagrams having the requested number of loops,
(ii) calculates the Hikami-boxes, and finally (iii) performs the integration over the cooperon and diffuson momenta.
The first two stages of the program are universal, i.e., can be readily used for other calculations in a
diagrammatic approach.
The program significantly facilitates the usage of the diagrammatics, especially in the spinful case.

The first part of the paper is not specific to the problem of anisotropic conductivity.
In Sec.~\ref{sec:Ham} we define the model which we use in our calculation, then we introduce disorder-averaged Green
functions (see Sec.~\ref{sec:GF}) and derive the Kubo-Greenwood formula in the Keldysh technique (see Sec.~\ref{sec:Kubo}).
Then we derive the loop expansion for the diagrammatic technique in Sec.~\ref{sec:le} and derive expressions for diffusons (cooperons) in Sec.~\ref{sec:CD}.
The second part of the paper starts with Sec.~\ref{sec:ZLA} where we derive the incoherent SOI-correction to the conductivity tensor. We then proceed with the
contribution from the weak-localization diagrams (which remains isotropic at zero frequency) in Sec.~\ref{sec:WL}. The results for the anisotropic transport in
2D and quasi-1D geometries \emph{in the presence of the time-reversal invariance} are described in Sec.~\ref{sec:twoLoop} for the case of zero frequency
$\omega=0$ and in Sec.~\ref{sec:ff} for $\omega\ne0$. Finally, in Sec.~\ref{sec:brokenTR} we give an example of how the effect of SOI-induced anisotropy could be enhanced
by the time-reversal symmetry breaking terms.

For convenience we summarize some often used notations in Tab.~\ref{tab:notations} at the end of the paper.

\begin{table}[f]
  \centering
  \begin{tabular}{|c|c|}\hline
Symbol & Its definition \tabularnewline\hline\hline
$d$& spatial dimension \tabularnewline\hline
$\sqrt i$&$(1+i)/\sqrt2$ \tabularnewline\hline
$C$ and $D$ & see~\eqref{obVyDC} and~\eqref{defXabDC} \tabularnewline\hline
$\GR$ and $\GA$ & see~\eqref{grSOI}; $\GA=\GR^\dag$ \tabularnewline\hline
$l$ &\begin{minipage}{.7\columnwidth} mean free path of an electron between subsequent elastic scattering off impurities\end{minipage}\tabularnewline\hline
$L_\phi$ &\begin{minipage}{.7\columnwidth} electron (orbital) dephasing length due to inelastic scattering\end{minipage}\tabularnewline\hline
$\tau_\phi^{-1}=\vf/L_\phi$ &\begin{minipage}{.7\columnwidth} dephasing rate due to inelastic scattering\end{minipage}\tabularnewline\hline
$\mu$ &(temperature-dependent) chemical potential\tabularnewline\hline
$\pf=\sqrt{2m\mu}$ &\begin{minipage}{.7\columnwidth} Fermi momentum\end{minipage}\tabularnewline\hline
$\vf=\pf/m$ &\begin{minipage}{.7\columnwidth} Fermi velocity\end{minipage}\tabularnewline\hline
$\nu$, DoS&density of states\tabularnewline\hline
$\Re$ and $\Im$& real and imaginary part\tabularnewline\hline
$\sigma_{\mathrm{is}}$ and $\sigma_{\mathrm{an}}$ & see~\eqref{defSan} and~\eqref{defSanInv}\tabularnewline\hline
$S$ & see~\eqref{pervoryadDSi} \tabularnewline\hline
$\sigma_0$ & $2\times2$ unity matrix \tabularnewline\hline
$\sigma_1$, $\sigma_2$, and $\sigma_3$ & Pauli matrices \tabularnewline\hline
$\xi_p$& see~\eqref{agf}\tabularnewline\hline
$\pf$, $x_a$, and $x_b$ & see~\eqref{dRDa} \tabularnewline\hline
$x$ and $\delta$ & see~\eqref{dRDaAlt} \tabularnewline\hline
CD&cooperon or diffuson\tabularnewline\hline
CS&coordinate system\tabularnewline\hline
DM&density matrix\tabularnewline\hline
GF&Green function\tabularnewline\hline
GFB&Green functions box\tabularnewline\hline
HB&Hikami box\tabularnewline\hline
lhs&left hand side\tabularnewline\hline
rhs&right hand side\tabularnewline\hline
SOI&spin-orbit interaction\tabularnewline\hline
VW&Vollhardt-Wölfle\tabularnewline\hline
WL&weak localization\tabularnewline\hline
ZLA&zero-loop approximation\tabularnewline\hline
  \end{tabular}
  \caption{Some often used notations and abbreviations.\label{tab:notations}}
\end{table}

\section{The Hamiltonian\label{sec:Ham}}
Rashba and Dresselhaus spin-orbit interaction terms modify the Hamiltonian as follows:
\begin{equation}\label{hamrash}
{\hat H}'=\frac{{\hat p}^2}{2m}+{\hat V}_s'+U'(\vec r),\quad V_s'=aV_R'+bV_D',
\end{equation}
where $\hat{\vec p}$ denotes momentum operator, $a$ and $b$ are Rashba and Dresselhaus amplitudes, and $U'(\vec r)$ is the disorder potential created by impurities or defects randomly
placed in the sample.  We assume that $U'(\vec r)$ is uncorrelated:
\begin{equation}\label{iznBes}
\overline{U'(\vec r)U'(\vec r')}=\frac{\hbar^2}{2\pi\nu\tau}\delta(\vec r-\vec r'),
\end{equation}
where $\tau$ is the mean time between collisions of an electron off impurities, $\nu$ is the density of states (DoS) at the Fermi level,
and the over-bar indicates average over the different disorder configurations.  The Rashba SOI is invariant under arbitrary rotation in the
$(x,y)$-plane:
\begin{equation}\begin{split}
V_R'=\hat{\vec z}\cdot\left[\boldsymbol\sigma\times\vec p\right]\equiv(\hat{\vec z},\boldsymbol\sigma,\vec p)=\sigma_1{\hat p}_y-\sigma_2{\hat p}_x\\
=(\hat{\vec z},R^z_\phi\boldsymbol\sigma,R^z_\phi\vec p)=\begin{pmatrix}0&p_y+ip_x\cr p_y-ip_x&0\end{pmatrix},\quad\forall\phi,
\end{split}\end{equation}
where $\boldsymbol\sigma=(\sigma_1,\sigma_2,\sigma_3)$ is composed of Pauli matrices, and $R^z_\phi$ denotes $3\times3$ matrix describing rotation by an angle
$\phi$ around the $z$-axis. The Dresselhaus SOI term can be written as
\begin{equation}\begin{split}
V_D'=(\hat{\vec z},C\boldsymbol\sigma,\vec p)=\sigma_1{\hat p}_x-\sigma_2{\hat p}_y,\\
C=R^z_{-\pi/2}R^y_\pi=\begin{pmatrix}0 & -1 &  0 \cr  -1 & 0 & 0 \cr 0 & 0 & -1\end{pmatrix}.
\end{split}\end{equation}
In the coordinate system (CS), rotated by an arbitrary angle $\phi$ around the $z$-axis with respect to the initial CS, the SOI part of the Hamiltonian is transformed into
\begin{equation}\label{COB}
V_s(\vec p,\boldsymbol\sigma)=V_s'(R^z_\phi\vec p,R^z_\phi\boldsymbol\sigma)=
(\hat{\vec z},a\boldsymbol\sigma+bR^z_{-\phi}CR^z_\phi\boldsymbol\sigma,\vec p).
\end{equation}
In case when $\phi=-\pi/4$, $V_D$ becomes similar to Rashba-SOI\cite{note:altPauliMatr}, so that the SOI term can be written as
\begin{equation}\begin{split}
R^z_{-\pi/4}CR^z_{\pi/4}&=\begin{pmatrix}-1&0&0\cr0&1&0\cr0&0&-1\end{pmatrix},\\ \label{neuSOI}
V_s\equiv V_s(\vec p,\boldsymbol\sigma)&=(a-b)\sigma_1\hat p_y-(a+b)\sigma_2\hat p_x={\hat s}{\hat\Delta}/2,
\end{split}\end{equation}
where the SOI-induced spectrum-splitting ${\hat\Delta}$ and the helicity (spirality\cite{Edelstein:90}) operators are defined as
\begin{equation}  \begin{split} \label{spectrumSplitting}
{\hat\Delta}=&2\sqrt{{\hat p}_x^2(a+b)^2+{\hat p}_y^2(a-b)^2},\\
{\hat s}=&2\left[(a-b){\hat p}_y\sigma_1-(a+b){\hat p}_x\sigma_2\right]{\hat\Delta}^{-1},\quad {\hat s}^2=\mathds1.
  \end{split} \end{equation}
The original disorder-free Hamiltonian in the rotated CS may be written in the form   ${\hat p}^2/(2m)+{\hat s}{\hat\Delta}/2$; it possesses the following eigensystem:
\begin{eqnarray}
|\vec p,s\rangle&=&\begin{pmatrix}\frac{is\Delta_{\vec p}/2}{(a+b)p_x+i(a-b)p_y}\\1\end{pmatrix}|\vec p\rangle,\quad
s=\pm1,\\
E_{\vec p,s}&=&\langle\vec p,s|{\hat H}_0|\vec p,s\rangle=\frac{p^2}{2m}+\frac{s\Delta_{\vec p}}2,\\
\label{spektr} \Delta_{\vec p}&=&\langle\vec p,s|{\hat\Delta}|\vec p,s\rangle=2\sqrt{p_x^2(a+b)^2+p_y^2(a-b)^2}.
\end{eqnarray}
We see that the simultaneous presence of Rashba and Dresselhaus SOI leads to an anisotropy of the energy spectrum.\cite{JC:2006,anisotrCond} On the other hand,~\eqref{spektr} is
symmetric with respect to the exchange $a\leftrightarrow b$; conversely, the same is true for SOI-induced corrections to the conductivity tensor.
We note that helicity is invariant with respect to the time reversal:
\begin{equation}
\sigma_2s_{-\vec p}^T\sigma_2=s_{\vec p}\equiv\langle \vec p|\hat s|\vec p\rangle.\label{invSpi}
\end{equation}
In the rest of the paper we perform  calculations in  the coordinate system, rotated by $\pi/4$ in the $xy$ plane, where the (unperturbed) Hamiltonian is given by
\begin{equation}\label{RDrotH}
{\hat H}=\frac{{\hat p}^2}{2m}+\frac{{\hat s}{\hat\Delta}}2+U(\vec r),
\end{equation}
where $U(\vec r)=U'(R^z_{\pi/4}\vec r)$ is the disorder potential in the rotated coordinate system.
Our assumption~\eqref{iznBes} that $U'(\vec r)$ is a $\delta$-correlated random disorder potential is inherited by the disorder potential in the rotated coordinate system:
\begin{equation}
\overline{U(\vec r)U(\vec r')}=\frac{\hbar^2}{2\pi\nu\tau}\delta(\vec r-\vec r').\label{fiMod}
\end{equation}
The Hamiltonian \eqref{RDrotH} defines the velocity operator $\hat{\vec v}$ together with the ``fictitious'' vector potential $\vec{\tilde A}$:
\begin{equation} \label{hamlo}\begin{split}
\hat{\vec v}=&\frac i\hbar[\hat H,\vec r]=\frac{\hat{\vec p}}m
-\frac e{mc}\vec{\tilde A},\\
\vec{\tilde A}=&\frac{mc}e\left[(a+b)\sigma_2,(b-a)\sigma_1,0\right],
\end{split}\end{equation}
so that
\begin{equation}
\frac{{\hat p}^2}{2m}+\frac{{\hat s}{\hat\Delta}}2=\frac{m{\hat v}^2}2-m(a^2+b^2).\label{eHtS}
\end{equation}
The strength of the SOI can be characterized by dimensionless Rashba and Dresselhaus amplitudes introduced as follows:
\begin{equation}\label{dRDa}
x_a=2\pf a\tau/\hbar,\quad x_b=2\pf b\tau/\hbar,\quad\pf=\sqrt{2m\mu},
\end{equation}
where $\mu$ is the (temperature-dependent) chemical potential.
Alternatively, the SOI can be characterized by another set of two dimensionless parameters:
\begin{equation}
x=\sqrt{x_a^2+x_b^2},\quad\delta=\frac{2ab}{a^2+b^2},\quad -1\le\delta\le1,    \label{dRDaAlt}
\end{equation}
which characterizes the ``total'' SOI amplitude and the anisotropy of the energy spectrum \eqref{spektr} correspondingly.
The choice between two parameter sets~\eqref{dRDa} and~\eqref{dRDaAlt} becomes important when we have to expand expressions in Taylor series.
On the example of the spectrum splitting $\Delta_{\vec p}$ defined in~\eqref{spektr} we see the advantage of the choice~\eqref{dRDaAlt}:
while the Taylor expansion of $\Delta_{\vec p}$ in powers of $(x,\delta)$ is uniform, its expansion in the
parameters~\eqref{dRDa} is non-uniform: the expansion depends on the fact if one expands subsequently in $x_a,x_b$ or in $x_b,x_a$.

\section{Averaged Green function in the self-consistent Born approximation\label{sec:GF}}
In our calculations, we use the electron-gas model of the Fermi liquid\cite{Abrikosov}.
In the absence of SOI and applied electric field, disorder-averaged Green functions are obtained from the self-consistent Born approximation~\cite{AGD}:
\begin{equation}\label{agf}
\grae(\vec p)=\left[E-\xi_{\vec p}\pm\frac i{2\tau}\right]^{-1},\quad\hbar\xi_{\vec p}=\frac{p^2}{2m}-\mu,
\end{equation}
where  $\tau$ has small ($\sim E\hbar/\mu$) dependence on frequency $E$.
The presence of the SOI changes the expression for Green function (GF) from~\eqref{agf} into
\begin{equation}\label{grSOI}\begin{split}
\GRE(\vec p)=&\gre(\vec p)\sum_{n\ge0}\left[V_s(\vec p)\gre(\vec p)/\hbar\right]^n\\
=&\left\{\sigma_0\left[\gre(\vec p)\right]^{-1}-\frac{s_{\vec p}\Delta_{\vec p}}{2\hbar}\right\}^{-1}\\
=&\frac12\left[\left(\sigma_0+s_{\vec p}\right)g_r^{E-}(\vec p)+\left(\sigma_0-s_{\vec p}\right)g_r^{E+}(\vec p)\right],
\end{split}\end{equation}
where $\sigma_0$ is the $2\times2$ unity matrix, and
\begin{equation}
g_r^{E\pm}(\vec p)=\left\{\left[\gre(\vec p)\right]^{-1}\pm\frac{{\Delta_{\vec p}}}{2\hbar}\right\}^{-1}=\left[g_a^{E\pm}(\vec p)\right]^*.\label{grapm}
\end{equation}
From \eqref{grSOI} we see that averaged GF for the Hamiltonian \eqref{RDrotH} can be obtained from the GF of the disorder-free Hamiltonian
by substituting infinitesimal ``epsilon'' in the denominator with $(2\tau)^{-1}$.
Thus for energies close to $\mu$, the averaged GFs are very different from GFs of the disorder-free system; the latter are strongly modified due to the disorder.
In fact, Eq.~\eqref{agf} is the result of the summation of an infinite perturbation series.

We calculate the universal contribution~\eqref{KuboFormula} to the conductivity tensor (see Sec.~\ref{sec:universal} below); the corresponding momentum integrals converge in the vicinity of
the Fermi level, so that the momentum arguments of all Green functions are close to $\pf$.
The assumption $p\approx\pf$ simplifies the SOI-term $V_s$ in the GF-expression~\eqref{grSOI}. In the zeroth order (in powers of $\hbar\xi_{\vec p}/\mu\ll1$)
\begin{equation}\label{genCubic}
V_s(\vec p)\approx\pf\left[(a+b)\sigma_1\sin\phi-(a-b)\sigma_2\cos\phi\right],
\end{equation}
where $\phi$ is the angular coordinate of $\vec p$. This approximation
is sufficient for the calculation of the weak localization and two-loop correction in Sec.~\ref{sec:WL} and~\ref{sec:twoLoop}.
However, in the calculation of the zero-loop contribution (see Sec.~\ref{sec:ZLA}), higher accuracy is required:
\begin{equation}\begin{split}
V_s(\vec p)\approx&\pf\left(1+\frac{\hbar\xi_{\vec p}}{2\mu}\right)\times\\
\times&\left[(a+b)\sigma_1\sin\phi-(a-b)\sigma_2\cos\phi\right].
\end{split}\end{equation}

\section{(Non)universal contributions to the conductivity\label{sec:universal}}
We calculate the universal (i.e., independent of the details of the energy spectrum far from the Fermi level) corrections to the conductivity tensor.  The latter quantity is derived in
linear response in the applied electric field (see Sec.~\ref{sec:Kubo}).
In the diagrammatic approach, the SOI-induced correction to the conductivity can be graphically represented as a sum of diagrams.
A contribution of an individual diagram  is initially expressed as an integral in both frequency and momentum %space
over the combination of Green functions and the distribution function.  We call such an integral \emph{universal} if its leading contribution comes from the part of the
integration space, where all momentum and frequency arguments of GFs in the integrand are close to the Fermi level.  In the momentum space this means $|p-\pf|\lesssim\hbar/l$; in the
frequency space $\hbar|E|\lesssim T$.  ($\pf$ is the Fermi momentum, and $T$ is the temperature in equilibrium or effective temperature in a non-equilibrium case.)  Then the integration
in momentum space can be performed assuming constant averaged DoS $\nu$ and approximating $\ipds\approx\nu\ixi$ where in 2D $\nu=m/(2\pi\hbar)$.
According to the Fermi liquid theory, only electrons with energies near the Fermi level behave like free electron gas so that the effect of the interaction between electrons can be
disregarded.
Thus only the universal corrections are expected to give reasonable physical results.
The non-universal contributions (i) cannot be reliably calculated and (ii) cannot cancel universal contributions.
In the diagrammatics, arbitrary universal contributions to Hikami boxes can be calculated.
Unfortunately, this is not true for non-universal corrections:
some of them can be considered within the diagrammatics\cite{note:NUC}, others are too complicated to be calculated.
The impossibility to take into account all non-universal contributions may leave an impression  of imperfection of the diagrammatic technique.
However, one should note that the situation is not better in the non-linear $\sigma$-model,\cite{Kamenev2}
where all approximations we use in the diagrammatics are required as well.

\section{Non-equilibrium Kubo formula in Keldysh technique\label{sec:Kubo}}
The (mean) current density in a system, characterized by a one-particle density matrix (DM) $\hat\rho$ is given by
\begin{equation}\label{tokFQ}
\vec j(t)=\Sp\left[{\hat\rho}(t){\hat{\vec j}}\right],\quad\hat{\vec j}=e\hat{\vec v},
\end{equation}
where $\hat{\vec j}$ and $\hat{\vec v}$ denote current and velocity operators. We proceed with calculations in momentum representation:
\begin{equation}\label{tokPKop}\begin{split}
\vec j(t)=\ipd\ippd\Sp_{\mathrm{spin}}\left[\langle\vec p|\hat\rho(t)|\vec p'\rangle
\langle\vec p'|\hat{\vec j}|\vec p\rangle\right],\\
\langle\vec p'|\hat{\vec j}|\vec p\rangle=\delta(\vec p-\vec p')e\frac{\partial H(\vec p,\vec r)}{\partial\vec p},\quad
\hat H={\hat H}_0+\delta\hat V,
\end{split}\end{equation}
where ${\hat H}_0$ is the unperturbed Hamiltonian, and the perturbation term $\delta\hat V$ describes the applied electric field, see~\eqref{perturbation}
below. It is convenient to express the DM in terms of GFs (see \S2.1 from Ref.~\onlinecite{Kadanoff}).
\bea
\langle\lambda|\hat\rho(t)|\lambda'\rangle=\langle{\hat\psi}^\dag(\lambda;t)\hat\psi(\lambda';t)\rangle
=-i\lim_{t'\to t}\langle\lambda'|{\hat G}^<(t',t)|\lambda\rangle=\nonumber\\
=-\frac i2\lim_{t'\to t}\langle\lambda'|\left[\hGK-\left(\hGR-\hGA\right)\right](t',t)|\lambda\rangle=
-\int_{-\infty}^\infty\frac{\ud\omega}{2\pi}e^{-i\omega t}\nonumber\\
\times\frac i2\iE\langle\lambda'|\left[\hGK-\left(\hGR-\hGA\right)\right](E,E-\omega)|\lambda\rangle,\quad
\label{perGF}\eea
where $\lambda\equiv(\vec p,s)$.
We assume that
the unperturbed DM~${\hat\rho}^{(0)}$ is stationary (though not necessary equilibrium) and is characterized by energy distribution function $f_E$.
Thus, the zero-order DM is time-independent, and the zero-order GFs are homogeneous in time.

The perturbation [see~\eqref{perturbation} below] affects both $\hat\rho(t)$ and $\hat{\vec j}$ in~\eqref{tokFQ}.
We call the correction to $\hat{\vec j}$  ``diamagnetic part of the current operator'' ${\hat{\vec j}}_D$;
the unperturbed part of~\eqref{tokPKop} we call ``normal part of the current operator'' ${\hat{\vec j}}_N$:
\begin{equation}\label{jND}\begin{split}
\langle\vec p'|\hat{\vec j}_D|\vec p\rangle=&\delta(\vec p-\vec p')e\frac{\partial\delta V}{\partial\vec p},\\
\langle\vec p'|{\hat{\vec j}}_N|\vec p\rangle=&\delta(\vec p-\vec p')e\frac{\partial H_0}{\partial\vec p}.
\end{split}\end{equation}
When the system is perturbed by external electric field $\vec E=-\frac1c\frac{\partial\vec A}{\partial t}$, the perturbation operator is given by
\begin{equation}\label{perturbation}\begin{split}
\langle&\vec p'|\delta\hat V|\vec p\rangle
=\delta(\vec p-\vec p')\frac1{2m}
\left\{\left[\vec p-\frac ec\left(\vec A+\vec{\tilde A}\,\right)\right]^2\right.\\
-&\left.\left[\vec p-\frac ec\vec{\tilde A}\right]^2\right\}\approx
-\frac e{mc}\vec A\left[\vec p-\frac ec\vec{\tilde A}\right]\delta(\vec p-\vec p')\\
=&-\frac1c\vec A\langle\vec p'|{\hat{\vec j}}_N|\vec p\rangle,\quad
\langle\vec p'|{\hat{\vec j}}_D|\vec p\rangle=-\frac{e^2}{mc}\delta(\vec p-\vec p')\vec A.
\end{split}\end{equation}
Let us denote the Keldysh-contour time ordered GF as $\hat G$.
The applied electric field affects $\hat G$; the first-order correction is given by
\begin{equation}\delta^{(1)}\hat G(E,E-\omega)={\hat G}^E\left[-\frac e{c\hbar}\hat{\vec v}\vec A_\omega\right]{\hat G}^{E-\omega},\label{ktoGF}\end{equation}
Expressing~\eqref{ktoGF} in a usual $2\times2$-matrix form\cite{Rammer}, we get perturbation expressions for GFs in~\eqref{perGF}:
\begin{equation}\begin{split}
\delta^{(1)}&(\hGR-\hGA)(E,E-\omega)\\
&=-\frac{eA_\omega^\beta}{c\hbar}\left[\hGRE{\hat v}_\beta\hGREw-\hGAE{\hat v}_\beta\hGAEw\right],
\end{split}\end{equation}
and
\begin{equation}\begin{split}\hspace{-1ex}
\delta^{(1)}&\hGK(E,E-\omega)=-\sum_{\beta=1}^2\frac{eA_\omega^\beta}{c\hbar}\left[\left(h_E-h_{E-\omega}\right)\hGRE{\hat v}_\beta\hGAEw\right.\\
+&\left.h_{E-\omega}\hGRE{\hat v}_\beta\hGREw-h_E\hGAE{\hat v}_\beta\hGAEw\right],\quad h_E=1-2f_E.
\end{split}\end{equation}
In equilibrium
\begin{equation}
h_E=\begin{cases} 0,&\ E\hbar<-\mu\\
\tanh\frac{E\hbar}{2T},&\ E\hbar\ge-\mu\end{cases},
\end{equation}
where $T$ is the temperature (measured in energy units).
We split the DM into ``normal'' and ``diamagnetic'' parts [like we did with the current operator in~\eqref{jND}]:
\begin{equation}\label{rhoND}\begin{split}
\langle\lambda|&\delta\hat\rho_N(\omega)|\lambda'\rangle=\frac i2\sum_{\beta=1}^2\frac{eA_\omega^\beta}{c\hbar}\times\\ \times&\iE
\langle\lambda'|\left[\left(h_E-h_{E-\omega}\right)\hGRE{\hat v}_\beta\hGAEw\right]|\lambda\rangle,
\end{split}\end{equation}
\begin{equation}\label{rhoDiam}\begin{split}
\langle\lambda|&\delta\hat\rho_D(\omega)|\lambda'\rangle=
\\ &=\frac i2\sum_{\beta=1}^2\frac{eA_\omega^\beta}{c\hbar}\iE
\langle\lambda'|\left[(h_{E-\omega}-1)\hGRE{\hat v}_\beta\hGREw\right.\\
&-\left.(h_E-1)\hGAE{\hat v}_\beta\hGAEw\right]|\lambda\rangle.
\end{split}\end{equation}
Then we rewrite~\eqref{tokFQ} in the frequency space, substituting
$\hat\rho(\omega)={\hat\rho}^{(0)}+\delta\hat\rho_N(\omega)+\delta\hat\rho_D(\omega)$:
\begin{equation}\label{SpND}\begin{split}
\vec j(\omega)=\Sp\left[\left(
{\hat\rho}^{(0)}+\delta{\hat\rho}_N(\omega)+\delta{\hat\rho}_D(\omega)\right)
\left({\hat{\vec j}}_N+{\hat{\vec j}}_D\right)\right]\\ \approx
\Sp\left[{\hat\rho}^{(0)}{\hat{\vec j}}_N+{\hat\rho}^{(0)}{\hat{\vec j}}_D+
\delta{\hat\rho}_N(\omega){\hat{\vec j}}_N+\delta{\hat\rho}_N(\omega){\hat{\vec j}}_D\right],
\end{split}\end{equation}
where we neglected non-linear (in the perturbation) terms.
Using Eqs.~\eqref{perGF}, \eqref{jND}, and~\eqref{rhoND} we can calculate~\eqref{SpND} in the momentum representation.
From
\begin{equation}\label{CancTrick}\begin{split}\hspace{-2ex}%
\Sp\left[v_i\hGRE v_j\hGRE\right]=\frac i\hbar&\Sp\left\{\left(r_i[\hGRE]^{-1}-[\hGRE]^{-1}r_i\right)\hGRE v_j\hGRE\right\} \\
=\frac i\hbar\Sp([r_i,v_j]\hGRE)&=\frac i\hbar\Sp\left(\left[r_i,\frac{{\hat p}_j}m\right]\hGRE\right)=-\delta_{ij}\Sp\hGRE
\end{split}\end{equation}
we conclude that
\begin{equation}
\text{at }\omega=0\quad \Sp\left[\delta{\hat\rho}_D\hat{\vec j}_N+\rho^{(0)}\delta{\hat{\vec j}}_D\right]=0.\label{diacan}
\end{equation}

Both equilibrium\cite{AmbegaokarEckern} and non-equilibrium\cite{PRLoctMMII} contributions to the persistent current in a mesoscopic ring are given
by~$\Sp\left[{\hat\rho}^{(0)}{\hat{\vec j}}_N\right]$, while~$\Sp\left[\delta{\hat\rho}_N(\omega){\hat{\vec j}}_N\right]$ is the linear response
to the applied electric field.

In the rest of this section we assume that $\vec A$ is directed along the $\beta$-axis, and we measure the charge current in the $\alpha$-direction.
Then (for arbitrary energy distribution~$f_E$)
\begin{equation}\label{KuboFormula}
  \begin{split}
    \sigma_{\alpha\beta}(\omega)&=\sigma^{\mathrm N}_{\alpha\beta}(\omega)+\sigma^{\mathrm D}_{\alpha\beta}(\omega),\quad
\sigma^{\mathrm N}_{\alpha\beta}(\omega)\gg\sigma^{\mathrm D}_{\alpha\beta}(\omega),\\
\hspace{-1ex}    \sigma^{\mathrm N}_{\alpha\beta}(\omega)=&\frac c{i\omega A_\omega}\Sp\left[\delta\hat\rho_N(\omega){\hat j}^{\,\alpha}_N\right]=
 \frac{e^2}h\Sp\left[{\hat v}_\alpha{\hat G}_{\mathrm R}^E{\hat v}_\beta{\hat G}_{\mathrm A}^{E-\omega}\right],
 \end{split}\end{equation}
where we assumed that the (momentum) trace is $E$-independent.
(This assumption is valid for all universal quantities except for the Drude conductivity, see Appendix~\ref{app:leadingCond}).
The diamagnetic correction to the conductivity is given by
\begin{equation} \label{correctionsToKubo} \begin{split}
\sigma^{\mathrm D}_{\alpha\beta}(\omega)=\frac c{i\omega A_\omega}\Sp\left[\delta\hat\rho_D(\omega){\hat j}^{\,\alpha}_N-\delta\hat\rho_D(0){\hat j}^{\,\alpha}_N\right]\\
=\frac{e^2}h\int_{-\infty}^\infty\frac{\ud E}\omega f_E\left\{
\Sp\left[{\hat v}_\alpha{\hat G}_{\mathrm A}^E{\hat v}_\beta\left({\hat G}_{\mathrm A}^{E-\omega}-{\hat G}_{\mathrm A}^E\right)\right] \right.\\\left.-
\Sp\left[{\hat v}_\alpha\left({\hat G}_{\mathrm R}^{E+\omega}-{\hat G}_{\mathrm R}^E\right){\hat v}_\beta{\hat G}_{\mathrm R}^E\right]\right\}.  
\end{split}\end{equation}
In the frequency integral $\iEs$ in~\eqref{correctionsToKubo}, large negative frequencies $E\sim-\mu/\hbar$ give important contribution to the result, so that the
$E$-dependence of $\tau$ in~\eqref{agf} cannot be neglected; this complicates the calculation of~\eqref{correctionsToKubo}.
The SOI-dependent part of~\eqref{correctionsToKubo} is non-universal (the momenta of GFs are not bounded in the vicinity of $\pf$).
Since~\eqref{correctionsToKubo} does not contain products of different (retarded and advanced) GFs, it cannot contain diffusons or cooperons; hence it is incoherent and cannot produce
corrections to the conductivity tensor having the same order in SOI, as our results below [see Eqs.~\eqref{leadingAnis}, \eqref{ani1D}, and~\eqref{piri}].
In what follows, we study the universal contribution~\eqref{KuboFormula} and do not calculate~\eqref{correctionsToKubo}.

An attempt to use the Kubo formula~\eqref{KuboFormula} for calculating the leading (Drude) conductivity contribution leads to divergences.
In fact, \eqref{KuboFormula} is valid only for calculating \emph{corrections}  (due to $\omega\ne0$, SOI, interaction, etc.) to the main (Drude) conductivity
value. See Appendix~\ref{app:leadingCond} for  details of calculating the Drude conductivity.

We derived the universal contribution to the conductivity tensor~\eqref{KuboFormula} for a general case of \emph{non-equilibrium stationary} distribution function.
The result~\eqref{KuboFormula} is the same as the one derived for the equilibrium case~\cite{JRammerQTT}.
Thus, we see that \emph{corrections} to the conductivity are independent of the distribution function~$f_E$.
Note that this is not true for the leading (Drude) conductivity~\eqref{inhCond}, which does depend on~$f_E$.

In what follows, we always perform calculations in the rotated coordinate system, where the spin-orbit part of the Hamiltonian is given by~\eqref{neuSOI}.
In this coordinate system, the conductivity tensor is diagonal in all considered geometries; its anisotropic part is proportional to $\sigma_3$.
[See the discussion after~\eqref{Aexpr}.]
We denote the isotropic and anisotropic parts of the conductivity tensor $\sigma$ with symbols $\sigma_{\mathrm{is}}$ and $\sigma_{\mathrm{an}}$:
\begin{equation}\label{defSan}
  \sigma=\sigma_{\mathrm{is}}\sigma_0+\sigma_{\mathrm{an}}\sigma_3.
\end{equation}
In an arbitrary coordinate system, the (an)isotropic properties of a 2D symmetric tensor can be characterized by two non-negative scalars
$\sigma_{\mathrm{is}}$ and $|\sigma_{\mathrm{an}}|$ --  isotropic and anisotropic amplitudes defined by
\begin{equation} \label{defSanInv} \begin{split}
    \sigma_{\mathrm{is}}=&\frac12\Sp\sigma,\quad |\sigma_{\mathrm{an}}|=\sqrt{\Sp\left[\left(\sigma-\sigma_0\sigma_{\mathrm{is}}\right)^2/2\right]} \\
=&\sqrt{\left\{\Sp\left[\sigma\frac{\sigma_1}2\right]\right\}^2+\left\{\Sp\left[\sigma\frac{\sigma_3}2\right]\right\}^2}.
  \end{split}\end{equation}
It is easy to check that both $\sigma_{\mathrm{is}}$ and $|\sigma_{\mathrm{an}}|$ are independent of the choice of the coordinate system.
Finally, we give explicit expressions for the anisotropic part~$\sigma_{\mathrm{an}}$ of the conductivity in the original and rotated by $\pi/4$ coordinate systems:
\begin{equation} \label{explAN} \begin{split}
    \text{in the original CS }\sigma_{\mathrm{an}}&=\sigma_{xy},\text{ and}\\
    \text{in the rotated CS }\sigma_{\mathrm{an}}&=\frac{\sigma_{xx}-\sigma_{yy}}2.\\
\end{split}\end{equation}

\section{The loop expansion\label{sec:le}}
It is convenient to represent the different contributions to the averaged conductivity in  graphical form as diagrams.
The simplest (bubble) diagram [see  Fig.~\ref{fZLA:a}]\cite{note:Drude} is produced by the Kubo formula~\eqref{KuboFormula} by substituting
${\hat G}_{\mathrm R}$ and ${\hat G}_{\mathrm A}$ with the averaged GFs $\GR$ [given by~\eqref{grSOI}] and $\GA=\left[\GR\right]^\dag$.
The bubble has neither diffuson nor cooperon lines.
One can proceed by connecting the retarded GF $\GR$ of a bubble with an advanced one $\GA$ by a cooperon or diffuson ladder  in all possible (two) ways.
Doing so we obtain diagrams depicted in Figures~\ref{fCondDiags} and~\ref{fZLA:b} containing one cooperon and one diffuson.
Adding more and more diffuson or cooperon lines in all possible ways, one obtains an infinite number of diagrams.
In this section we describe how the most important diagrams can be selected out of this infinity for further calculation.
\subsection{Two ways of drawing diagrams\label{sec:twoWays}}
\newcommand{\evA}{-e\vec v\vec A}\newcommand{\jcl}{\hat{\vec j}}%
\begin{figure}%
\subfigure[]{\label{fCondDiags:a}\begin{minipage}{.4\columnwidth}\resizebox{\textwidth}{!}{\input{cond1.ins.tex}}\end{minipage}}\hspace{.1\columnwidth}%
\subfigure[]{\label{fCondDiags:b}\begin{minipage}{.4\columnwidth}\resizebox{\textwidth}{!}{\input{cond1a.ins.tex}}\end{minipage}}\hspace{.1\columnwidth}%
\mycaption{Two representations of the weak localization diagram, cf. Fig.~4.8 from Ref.~\onlinecite{montambauxBook}.
We call Fig.~\ref{fCondDiags:a} ``ladder representation'', and Fig.~\ref{fCondDiags:b} -- ``coordinate representation''.\label{fCondDiags}}
\end{figure}
In Fig.~\ref{fCondDiags}, the same (weak localization) diagram is drawn in two equivalent representations:
on the lhs, the cooperon is drawn in the ``ladder'' form, [see the lhs of Fig.~\ref{fCoopDiff:b}] while on the rhs the ``coordinate'' form
[wavy line with two ends, see the rhs of Fig.~\ref{fCoopDiff:b}] is used.

The weak localization diagram is usually drawn in the lhs (ladder) -form (or as topologically-equivalent ``bubble with maximally anti-crossing
disorder-averaging lines'', cf. Fig.~4.8 from Ref.~\onlinecite{montambauxBook}). Below we use the rhs (coordinate)-form
[its advantages are discussed in Sec.~\ref{sec:defCD} below]; Fig.~\ref{fCoopDiff} gives a recipe how a
diagram can be transformed from one form into another and back.

A diagram in the coordinate representation consists of Green function boxes (bubbles, triangles, squares, pentagons, etc.) connected by wavy lines (cooperons
and/or diffusons). A vertex of a Green function box (GFB) may be occupied by a (i) observable operator, (ii) external field operator, or (iii) end of
a cooperon and/or diffuson line.

\subsection{Loops formed by cooperons and diffusons\label{sec:loops}}
There are two important momentum scales in the disorder averaging technique:
(i) the (characteristic) absolute value of the momentum argument in averaged GFs $p\sim\pf$, and (ii) $\hbar/l\ll\pf$.
($l$ is the mean free path of an electron between two subsequent elastic scatterings off impurities.)
Momentum integrals from products of GFs of the form
\begin{equation}  \ipd\prod_{i=1}^r\GR\left(\vec p-{\vec q}_i\right)\prod_{j=1}^a\GA\left(\vec p-{\vec q}_j\right)\end{equation}
usually converge within the interval $\pf-\hbar/l\lesssim p\lesssim\pf+\hbar/l$; hence $\hbar/l$ characterizes the deviation of momentum argument of an
averaged GF ~\eqref{agf} or~\eqref{grSOI} from $\pf$.
The assumption that ``large''  momentum $\pf$  is much larger than ``small''  momentum $\hbar/l$ is crucial for the disorder averaging technique, since
$(\pf l/\hbar)^{-1}\ll1$ is its main expansion parameter (see Sec.~\ref{sec:compareTwo} below).

The mean scattering free path $l$ is also a scale on which
averaged GFs \eqref{agf} and \eqref{grSOI} decay, e.g., in 3D $\GRA(\vec r-\vec r')\propto\exp[-|\vec r-\vec r'|/l]$.
We can interpret this saying that the length of a Green function line in our diagrams is $l$.
Within the disorder averaging technique we  cannot observe  effects on scales shorter than $l$, i.e., to say that the length
of a GF-line is $l$, is almost the same, as to say that this length is zero; thus we can consider a GF-line not as a line, but as a point.

Now if we draw some diagram in its ``coordinate representation'' (see Sec.~\ref{sec:twoWays}) and
squeeze all Green function lines into points, the result will contain only cooperon or diffuson (CD) lines forming a certain number of
loops.  For example, a bubble in Fig.~\ref{fZLA:a} has no loops (since it has no CD lines which
could form a loop); a weak localization (WL) diagram in Fig.~\ref{fCondDiags:b} has one loop, and all diagrams in Figs.~\ref{fdivtoPor},\ref{fOtherSL} have two loops.
\begin{figure}
\subfigure[]{\label{fZLA:a}%
\begin{minipage}{.4\columnwidth}\resizebox{\textwidth}{!}{\input{drude.ins.tex}}\end{minipage}}\qquad
\subfigure[]{\label{fZLA:b}%
\begin{minipage}{.4\columnwidth}\resizebox{\textwidth}{!}{\input{vertexRenorm.ins.tex}}\end{minipage}}\hspace{.07\columnwidth}%
\mycaption{Zero-loop diagrams: (a) the Drude bubble and  (b) the vertex renormalization.\label{fZLA}}
\end{figure}

The number of CD-loops is equal to the number of independent ``small'' momentum variables (which are $\sim\hbar/l$), or, in other words -- to the
number of integrals over the CD-momentum variables.

\subsection{Comparing two arbitrary diagrams\label{sec:compareTwo}}
Let us estimate two arbitrary diagrams for the same physical quantity.
An estimate for a GFB is $\propto\nu\tau^{h-1}$, where $h$ is the number of Green function lines composing the GFB.
Every CD line has a prefactor $(4\pi\nu\tau)^{-1}$ [see Sec.~\ref{sec:defCD} below]. We estimate the  DoS  as $\nu\sim m/(2\pi\hbar\lambdaslash^{d-2})$,
where $\lambdaslash\sim\hbar/\pf$, and $d$ is the spatial dimension ($d=2$ or $d=3$).
Let us denote $L_{1,2}$, $H_{1,2}$, $C_{1,2}$ the corresponding number of loops, GFBs, and CD-lines in the two considered diagrams; the
quantities $h_{1j}$ denote number of GF lines in the $j$th GFB of the first diagram, and $h_{2n}$ do the same for the
second diagram.

The calculation of diagrams is often much simpler in the diffusion approximation  -- i.e., assumption that $q^*l\ll\hbar$, 
where $q^*$ is the characteristic momentum of a CD line (i.e., ``small'' momentum variable).
[The validity of the diffusion approximation in our calculation arises from the assumption that $q^*l\sim x\hbar\ll\hbar$.]
Sometimes a GFB gains additional smallness of the order of $q^*l/\hbar\ll1$.
One has to calculate a GFB in order to reveal how much of these ``extra'' $q^*l/\hbar$ it has -- this is not uniquely defined by the number of loops.
In the following estimates we assume $q^*l\sim\hbar$ in order not to mix up  expansions in two different small parameters: $q^*l/\hbar\ll1$
and~$(\pf l/\hbar)^{-1}\ll1$. Then the relation between two different arbitrary diagrams is estimated as
\begin{equation}
\hspace{-3ex}\frac{\mathrm{1st\ diagram}}{\mathrm{2nd\ diagram}}\sim\frac%
{\int\prod_{i=1}^{L_1}\frac{\ud^dk_i}{(2\pi\hbar)^d}\left[\prod_{j=1}^{H_1}2\pi\nu\tau^{h_{1j}-1}\right]\left[\frac1{2\pi\nu\tau}\right]^{C_1}}%
{\int\prod_{l=1}^{L_2}\frac{\ud^dq_l}{(2\pi\hbar)^d}\left[\prod_{n=1}^{H_2}2\pi\nu\tau^{h_{2n}-1}\right]\left[\frac1{2\pi\nu\tau}\right]^{C_2}}.
\end{equation}
We use the fact that $L_i=C_i-H_i+1$ for $i=1,2$, and $\sum_{j=1}^{H_1}h_{1j}-\sum_{n=1}^{H_2}h_{2n}=2(C_1-C_2)$, so that
\begin{equation}\begin{split}\label{boPa}
&\frac{\mathrm{1st\ diagram}}{\mathrm{2nd\ diagram}}\sim
\left[\frac1{(2\pi)^d}\left(\frac\hbar{\pf l}\right)^{d-1}\right]^{L_1-L_2},\quad d>1.
\end{split}\end{equation}
As we discussed above, apart from $(\pf l/\hbar)^{-1}\gg1$, there is an additional small expansion parameter $q^*l/\hbar\ll1$; so the total number of expansion parameters is two.
(Later, in Sec.~\ref{sec:ff} the number of small parameters is three.)
The loop expansion predicts how large (small)  an arbitrary diagram is \emph{only} in powers of $(\pf l/\hbar)^{-1}$.

To conclude, the statement that ``every loop brings a smallness $(\pf l/\hbar)^{-1}$'' is known in mesoscopics; for the diagrams produced by the non-linear
$\sigma$-model\cite{KamenevAndreevNLSM} it is explained in Sec.~III.3.c of Ref.~\onlinecite{BelitzKirkpatrick}.  However, we are not aware of earlier articles,
where this statement is justified for diagrams within the usual disorder-averaging diagrammatic technique; this was the reason to include Sec.~\ref{sec:le} in this text.

\section{Zero-frequency cooperon and diffuson in the presence of SOI\label{sec:CD}}
Both cooperon and diffuson\cite{montambauxBook} can be graphically represented as a sum of ``ladders''; each ``ladder'' is given by retarded Green function
$\GR$ (bold line in our drawings) connected to the advanced one $\GA$ (bold line) with some number of disorder-averaging (dashed) lines.  An elementary building block
of every such ladder is made of one dashed line connecting $\GR$ with $\GA$. Below we discuss a convenient way of rearranging spin indices in every building block of the ladder.

\subsection{Separating spin indices}
One can write the expression for two GF-lines connected with a disorder-averaging line in different ways; we will write it either as
\begin{widetext}
\begin{equation}\label{eblo}
  \begin{split}
    \begin{minipage}[c]{.16\columnwidth}\resizebox{!}{6.5ex}{\input{di1.ins.tex}}\end{minipage}=\frac1{2\pi\nu\tau} \left[\GR(\vec p_1)\GR(\vec p_2)\right]_{\alpha\beta} \left[\GA(\vec p_1')\GA(\vec
      p_2')\right]_{\beta'\alpha'}\\
= \frac1{4\pi\nu\tau}\sum_{l=0}^3 \left[\GR(\vec p_1)\sigma_l\GA(\vec p_1')\right]_{\alpha\alpha'} \left[\GA(\vec
      p_2')\sigma_l\GR(\vec p_2)\right]_{\beta'\beta},
  \end{split}
\end{equation}
  or as
\begin{equation}\label{ebloC}
  \begin{split}
    \begin{minipage}[c]{.16\columnwidth}\resizebox{!}{6.5ex}{\input{co1.ins.tex}}\end{minipage}=\frac1{2\pi\nu\tau} \left[\GR(\vec p_1)\GR(\vec p_2)\right]_{\alpha\beta} \left[\GA(\vec p_1')\GA(\vec
      p_2')\right]_{\alpha'\beta'}\\
= \frac1{4\pi\nu\tau}\sum_{l=0}^3 \left[\GR(\vec p_1){\bar\sigma}_l^\dag G_{\mathrm A}^T(\vec p_1')\right]_{\alpha\alpha'}
    \left[G_{\mathrm A}^T(\vec p_2'){\bar\sigma}_l\GR(\vec p_2)\right]_{\beta'\beta},
  \end{split}
\end{equation}
where  the  identity is used:
\begin{equation}\label{borinoTozhdestvo}
2\delta_{s_1s_2}\delta_{s_3s_4}=\sum_{\alpha=0}^3\sigma^{s_1s_3}_\alpha{\sigma}^{s_4s_2}_\alpha=
\sum_{\alpha=0}^3{\bar\sigma}^{s_1s_3}_\alpha\left({\bar\sigma}_\alpha^\dagger\right)^{s_4s_2},\quad
{\bar\sigma}_\alpha\equiv\sigma_2\sigma_\alpha.
\end{equation}
Note that, differently from the first lines of \eqref{eblo} and~\eqref{ebloC}, square brackets in their second lines
contain spin and momenta variables belonging only to the pair of GFs from one side of the diagram (the lhs or the rhs from the central disorder-averaging line).
This ``separation of spin indices'' effectively makes the lhs of the diagrams in~\eqref{eblo} and in~\eqref{ebloC} independent from their rhs.

Every diagram with cooperons or diffusons contains an infinite number of elementary blocks \eqref{eblo} and/or~\eqref{ebloC}.
Below we will always separate spin variables in them according to~\eqref{eblo} or~\eqref{ebloC}.

\subsection{Defining cooperon and diffuson\label{sec:defCD}}
Now let us transform \emph{every} elementary building block in the diffuson series according to~\eqref{eblo}:
\begin{equation}\label{ids}    \begin{split}
\begin{minipage}[c]{.16\columnwidth}\resizebox{!}{6.5ex}{\input{di1.ins.tex}}\end{minipage}\ +\ \begin{minipage}[c]{.16\columnwidth}\resizebox{!}{6.5ex}{\input{di2.ins.tex}}\end{minipage}\ +\ \begin{minipage}[c]{.18\columnwidth}\resizebox{!}{6.5ex}{\input{di3.ins.tex}}\end{minipage}\ +\ \ldots\\
=\sum_{l,l'=0}^3D^{ll'}_{\vec p_1-\vec p_1'} \left[\GR(\vec p_1)\sigma_l\GA(\vec p_1')\right]_{\alpha\alpha'}\left[\GA(\vec p_2')\sigma_{l'}\GR(\vec p_2)\right]_{\beta'\beta},
\end{split}  \end{equation}
where $\vec p_1-\vec p_1'=\vec p_2-\vec p_2'$ and
\begin{equation}\label{defXabDC}
    D^{\alpha\beta}_{\vec q}=\frac1{4\pi\nu\tau}\left[\sum_{n\ge0}X_D^n\right]_{\alpha\beta}
    =\frac1{4\pi\nu\tau}\left[\mathds1-X_D(\vec q\,)\right]^{-1}_{\alpha\beta},\quad
    X_D^{\alpha\beta}(\vec q)=\frac1{4\pi\nu\tau}\int\frac{\ud^2p}{(2\pi\hbar)^2}\Sp_{\mathrm{spin}}[\sigma_\alpha\GRE(\vec
    p)\sigma_\beta\GAEw(\vec p-\vec q)],
\end{equation}
where $\Sp_{\mathrm{spin}}$ stands for the trace only in spin indices. So \eqref{eblo} helped us to convert the diffuson series into geometric series  that
we could sum.  Analogously we use~\eqref{ebloC} to transform the cooperon\cite{note:coopDef} series:
\begin{equation}\begin{split}\label{ics}
    \begin{minipage}[c]{.16\columnwidth}\resizebox{!}{6.5ex}{\input{co1.ins.tex}}\end{minipage}\ +\ \begin{minipage}[c]{.16\columnwidth}\resizebox{!}{6.5ex}{\input{co2.ins.tex}}\end{minipage}\ +\ 
    \begin{minipage}[c]{.18\columnwidth}\resizebox{!}{6.5ex}{\input{co3.ins.tex}}\end{minipage}\ +\ \ldots\\
=\sum_{l,l'=0}^3C^{ll'}_{\vec p_1+\vec p_2} \left[\GR(\vec p_1)\sigma_l\GA(\vec p_1')\right]_{\alpha\alpha'}\left[\GA(\vec p_2')\sigma_{l'}\GR(\vec p_2)\right]_{\beta'\beta},
\end{split}\end{equation}
where $\vec p_1+\vec p_1'=\vec p_2+\vec p_2'$ and
\begin{equation}\label{obVyDC}
C^{\alpha\beta}_{\vec q}=\frac1{4\pi\nu\tau}\left[\sum_{n\ge0}X_C^n\right]_{\alpha\beta}=\frac1{4\pi\nu\tau}\left[\mathds1-X_C(\vec q\,)\right]^{-1}_{\alpha\beta},\quad
X_C^{\alpha\beta}(\vec q)=\frac1{4\pi\nu\tau}\int\frac{\ud^2p}{(2\pi\hbar)^2} \Sp_{\mathrm{spin}}\left\{{\bar\sigma}_\alpha\GRE(\vec p){\bar\sigma}_\beta^\dag[\GAEw(\vec q-\vec p)]^T\right\}.
\end{equation}
\end{widetext}
From~\eqref{invSpi} it follows  that \emph{in a system with time reversal symmetry}
\begin{equation}
\sigma_2G_{\mathrm R/A}^T(-\vec p)\sigma_2=\GRA(\vec p),\label{timeRevarsalGF}
\end{equation}
so that
\begin{equation}
X_C^{\alpha\beta}(\vec q)=X_D^{\alpha\beta}(\vec q) \mathrm{\ and\ } C^{\alpha\beta}_{\vec q}=D^{\alpha\beta}_{\vec q}.\label{CeqD}
\end{equation}

The series~\eqref{ids} [or~\eqref{ics}] depend on four momentum and four spin variables;
without external four GF lines only $D^{ll'}_{\vec p_1-\vec p_2}$ [or $C^{ll'}_{\vec p_1+\vec p_2}$] is left, which is a function of two momentum and two spin variables.

We call the quantities $D^{\alpha\beta}_{\vec q}$ and $C^{\alpha\beta}_{\vec q}$ the diffuson and the cooperon, respectively.

Diagrammatically, $D^{\alpha\beta}_{\vec q}$ can be drawn in two ways. The lhs-diagram in Fig.~\ref{fCoopDiff:a} is more similar to the series
in~\eqref{ids}, while the rhs-diagram in Fig.~\ref{fCoopDiff:a} stresses the fact that the diffuson without external four GF lines has only two ends.
\begin{figure}
\subfigure[]{\label{fCoopDiff:a}%
\begin{minipage}{.45\columnwidth}\resizebox{\textwidth}{!}{\input{diffuson.ins.tex}}\end{minipage}}\quad
\subfigure[]{\label{fCoopDiff:b}%
\begin{minipage}{.45\columnwidth}\resizebox{\textwidth}{!}{\input{cooperon.ins.tex}}\end{minipage}}\hspace{.07\columnwidth}%
\mycaption{Diagrams for (a)  diffuson and (b) cooperon in two representations.  On the lhs, the diffuson (cooperon) is drawn as a ``ladder''; on the rhs it is drawn
  as a wavy line. (See also Sec.~\ref{sec:twoWays} and Fig.~\ref{fCondDiags}.)\label{fCoopDiff}}
\end{figure}

This rhs-diagram in Fig.~\ref{fCoopDiff:a} reflects better the spatial structure of the diffuson in the coordinate space.
From Eq.~\eqref{fiMod} one can see that
the distance between points 1 and 2 in Fig.~\ref{fCoopDiff} is zero, and the same
is true for the points 3 and 4. We merged these identical points in the rhs-diagram in Fig.~\ref{fCoopDiff:a}, so that 1=2 and 3=4.
A diffuson at frequency $\omega$ in coordinate space decays on the scale $\min(|L_\omega|,L_s)\gg l$:
\begin{equation}\label{Lw}\begin{split}
L_\omega=l/\sqrt{2i\omega\tau},\quad\sqrt i\equiv(1+i)/\sqrt2,\quad L_s=l/\sqrt{2x},\\
\omega\tau\ll\hbar,\quad x\ll1\Longrightarrow\min(|L_\omega|,L_s)\gg l.
\end{split}\end{equation}
Thus, we see that the distance (in coordinate space) between points 1 and 2 (or 3 and 4) in Fig.~\ref{fCoopDiff:a} is (within our accuracy) infinitesimal,
and this fact is graphically reflected by merging points 1 and 2 together (as well as points 3 and 4) in the rhs of Fig.~\ref{fCoopDiff:a}.

A similar reasoning is valid for the cooperon as well, see Fig.~\ref{fCoopDiff:b}.

\subsection{Explicit 2D-expressions for $q=0$\label{sec:zeroQ}}
The diffuson at zero momentum, $q=0$, can be calculated without assuming that the SOI amplitudes $x_{a,b}$ are small.
Using the fact  that
\bes
\GR^T(\vec p)+\GR^T(-\vec p)=\GR(\vec p)+\GR(-\vec p),
\ees
one obtains a sum rule:
\begin{equation}\label{Xsr} X_D^{22}=X_D^{00}-X_D^{11}+X_D^{33}.\end{equation}
For $q=0$, $X_D$ is a diagonal $4\times4$ matrix with elements
\begin{equation}\begin{split}\label{toVyddpqrn}
X_D^{00}=1,\ X_D^{11}=\frac{1+K}{1+(x_a+x_b)^2+K}\ ,\ X_D^{33}=\frac1K\ ,\\
K=\sqrt{[1+(x_a+x_b)^2][1+(x_a-x_b)^2]}.
\end{split}\end{equation}
The components of the diffuson at zero momentum are given by the diagonal matrix
\begin{equation}\label{exactDiffusonForZeroQ}\begin{split}
4\pi\nu\tau D_0&=\frac{2m\tau}\hbar D_0\\
=\mathrm{diag}&\left(\frac{L_\phi^2}{l^2},1+\frac{1+K}{(x_a+x_b)^2},1+\frac{1+K}{(x_a-x_b)^2},\frac K{K-1}\right),
\end{split}\end{equation}
where the electron (orbital) dephasing length $L_\phi$ (due to inelastic scattering) serves as a cut-off for the infinity,
and the first element $L_\phi^2/l^2$ does not contribute to physical quantities when the Vollhardt-Wölfle theorem holds (see Appendix~\ref{app:VW}).
The sum of the two diagrams in Fig.~\ref{fZLA} is equal to the diagram in Fig.~\ref{fZLA:a} with one velocity vertex renormalized:
\begin{equation}\label{renVelGraph}
\begin{minipage}{3ex}\includegraphics[height=6ex]{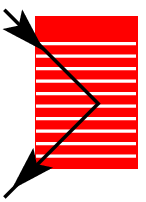}\end{minipage}=
\begin{minipage}{3ex}\includegraphics[height=6ex]{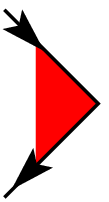}\end{minipage}+
\begin{minipage}{14.2ex}\includegraphics[height=6ex]{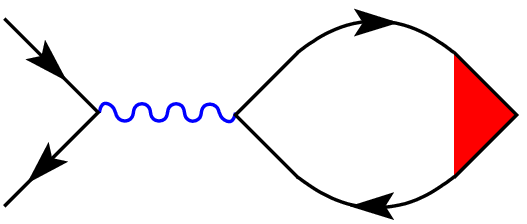}\end{minipage},
\end{equation}
which corresponds to the expression
\begin{equation}\label{renVel}
{\tilde v}_\alpha=
v_\alpha+\sum_{\gamma=1}^3\sigma_\gamma D_0^{\gamma\gamma}\Sp\left[\sigma_\gamma\GR(\vec p)v_\alpha\GA(\vec p)\right]=\frac{p_\alpha}m,
\end{equation}
where $D_0^{\gamma\gamma}$ are components of the zero-momentum diffuson given by~(\ref{exactDiffusonForZeroQ},\ref{toVyddpqrn}).

Since the vertex renormalization of this type occurs in every diagram for the conductivity, we take it everywhere into account by substituting the velocity operator
$v_\alpha$ with its renormalized value ${\tilde v}_\alpha=p_\alpha/m$; the only exception is the zero-loop contribution (calculated in Sec.~\ref{sec:ZLA}), which is
represented by the diagram in Fig.~\ref{fZLA:a} with \emph{only one} velocity vertex being renormalized.

\subsection{Explicit 2D-expressions for $q\ne0$\label{sec:dbnez}}
Because of the SOI, most components of the diffuson do not have a pole at zero momentum and frequency even if the dephasing effects are neglected; the diffuson gains a
non-zero ``mass'', see \eqref{exactDiffusonForZeroQ}. In case of pure-Rashba or pure-Dresselhaus SOI, this ``mass'' is quadratic in the SOI amplitude.  This still remains true in case
when one (Rashba or Dresselhaus) SOI amplitude is much smaller than the other one, so that the SOI-induced anisotropy of the energy spectrum \eqref{dRDaAlt} $\delta$ is a small
parameter. Consequently, the integrals in diffuson momenta $\vec k$ and $\vec q$ converge on the scale of $q\lesssim x\hbar/l$, and it is convenient to introduce dimensionless variables
\begin{equation}
\vec K\equiv l\vec k/x\hbar,\quad\vec Q\equiv l\vec q/x\hbar,\label{defBP}
\end{equation}
so that $K\lesssim1$ and $Q\lesssim1$. Like in Sec.~\ref{sec:dbnez}, the calculation shows that only the $(1,1)$-minor (i.e., $3\times3$ matrix block) of the $4\times4$
diffuson matrix is affected by the SOI. In other words, the upper line and the left column of the diffuson matrix are independent of SOI:
\begin{equation}
\begin{split}D^{00}_{\vec q}=\frac\hbar{2m\tau}\frac1{l^2q^2/2\hbar-i\omega\tau},\\ D_{\vec q}^{\alpha0}=D_{\vec q}^{0\alpha}=0,\quad\alpha=1\ldots3.\end{split}
\end{equation}
The element $D_{\vec q}^{00}$ gives no contribution to the conductivity in systems with time-reversal symmetry (see Appendix~\ref{app:VW}); it becomes
important, when this symmetry is broken, see Sec.~\ref{sec:brokenTR}. In the rest of this Section we reduce $4\times4$ matrices of $X_D$ and $D_{\vec q}$ to
corresponding (1,1)-minors.

To simplify the calculation, we further assume that $x\ll1$ (diffusion approximation). Then
the diffuson is obtained using \eqref{defXabDC} with $X_D$ given by
\begin{equation}X_D\approx\mathds1-x^2\left[Y^{(0)}_{\vec Q}-\delta Y^{(1,0)}_{\vec Q}\right],\end{equation}
where $\delta$ is defined in~\eqref{dRDaAlt}, and
\bea\label{XdRD}
Y^{(0)}_{\vec Q}&=&\frac{Q^2}2\mathds1+\frac12\begin{pmatrix}
1     &0           &-2iQ_x\\
0     &1           &-2iQ_y\\
2iQ_x &2iQ_y       & 2   \end{pmatrix},\\
Y^{(1,0)}_{\vec Q}&=&\frac12\begin{pmatrix}
-1     & 0           & iQ_x\\
 0     & 1           &-iQ_y\\
-iQ_x  & iQ_y        & 0   \end{pmatrix}.
\eea
Note that the above expression for $X_D$ is Hermitian and obeys the sum rule \eqref{Xsr}.

The resulting diffuson matrix $D_{\vec Q}$ has a denominator $\propto(\det Y^{(0)}_{\vec Q})^n$, where $n>0$ is an integer, and
\begin{equation}\label{detYZ}\begin{split}
8\det Y^{(0)}_{\vec Q}&=2+Q^2+Q^6\\
=&(Q^2+1)\left(Q^2-\frac{1-i\sqrt7}2\right)\left(Q^2-\frac{1+i\sqrt7}2\right).
\end{split}\end{equation}
The expression \eqref{detYZ} is independent of the direction of $\vec Q$,
so the same is true for the denominators of all diffuson components.
Consequently, an arbitrary diagram containing a diffuson line with a non-zero momentum (e.g., rhs of Fig.~\ref{fdivtoPor}) has denominators [consisting of
powers of $\det Y^{(0)}_{\vec K}$, $\det Y^{(0)}_{\vec K+\vec Q}$, and $\det Y^{(0)}_{\vec Q}$],
which are invariant with respect to two ``mirror reflections'':
(i) $(K_x\to -K_x,Q_x\to -Q_x)$, (ii) $(K_y\to -K_y,Q_y\to -Q_y)$, and (iii)
$(K_x\leftrightarrow K_y,Q_x\leftrightarrow Q_y)$.
[Note that the original Hamiltonian~\eqref{hamrash}, \eqref{RDrotH} does not possess any of these symmetries.]
These symmetries are used in the program\cite{theProgram} for reducing the size of the integrands.

Finally, we note that the easiest way to obtain the results of this section is to utilize the computer program\cite{theProgram}.

\section{The zero-loop contribution\label{sec:ZLA}}
In the zero-loop approximation (ZLA), the calculation can be performed without assuming that SOI amplitudes are small, $x_{a,b}\ll1$ (i.e., without assuming the validity of the diffusion approximation).
Only two diagrams (see Fig.~\ref{fZLA}) having zero loops contribute to the ZLA. As we discussed in Sec.~\ref{sec:zeroQ}, their sum is equal to the diagram in
Fig.~\ref{fZLA:a} with one velocity vertex substituted by its renormalized value~\eqref{renVel}:
\begin{equation}\label{obVydpsoi}\begin{split}
\sigma^{(0)}_{\alpha\beta}&-\delta_{\alpha\beta}\sigma_D=\frac{e^2}h\Sp\left[{\hat v}_\alpha\hGR\frac{{\hat p}_\beta}m\hGA-\sigma_0\frac{p_\alpha p_\beta}{m^2}\hgre\hgae\right]\\
=\frac{e^2}h&\ipd\frac{p_\alpha p_\beta}{m^2}\left[\gr^{E-}\ga^{E-}+\gr^{E+}\ga^{E+}-2\hgre\hgae\right]\\
&-\frac{e^2}{2h}\Sp\left[\left(\frac e{mc}{\tilde A}_\alpha\right)\frac{{\hat p}_\beta}m\hat s(\gr^{E-}\ga^{E-}-\gr^{E+}\ga^{E+})\right],
\end{split}\end{equation}
where $g_{r/a}^{E\pm}$ are defined in \eqref{grapm}, and we subtracted the SOI-independent Drude conductivity $\sigma_D$ \eqref{inhCond}.
The second line of~\eqref{obVydpsoi} gives
\begin{equation}\label{spPrToIm}\begin{split}
\ipd\frac{p_\alpha p_\beta}{m^2}\left(\gr^{E-}\ga^{E-}+\gr^{E+}\ga^{E+}-2\gr^E\ga^E\right)\\
=\frac\hbar{2\mu\tau}\left[(x_a^2+x_b^2)\sigma_0+x_ax_b\sigma_3\right],
\end{split}\end{equation}
where we used \eqref{genCubic}.
The rest of \eqref{obVydpsoi} equals
\begin{align}
\frac12&\Sp\left[\left(-\frac e{mc}{\tilde A}_\alpha\right)\frac{{\hat p}_\beta}m\hat s(\gr^{E-}\ga^{E-}-\gr^{E+}\ga^{E+})\right]\nonumber\\
=&\frac12\sum_{i=1}^2\left(\frac e{mc}\right)^2\Sp\left[{\tilde A}_\alpha {\tilde A}_i
\frac{2p_\beta p_i}{m\Delta_{\vec p}}\cdot(\gr^{E-}\ga^{E-}-\gr^{E+}\ga^{E+})\right]\nonumber\\
&=\frac\hbar{2\mu\tau}\left[-\frac{x_a^2+x_b^2}2\sigma_0-x_ax_b\sigma_3\right]. \label{okoRdvena}
\end{align}
We see that the anisotropic terms in \eqref{spPrToIm} and \eqref{okoRdvena} cancel each other so that the
charge conductivity \eqref{obVydpsoi} is proportional to the unity tensor\cite{anisotrCond}.
Thus, within the ZLA~$\sigma_{\mathrm{an}}=0$ and
\begin{equation}\label{ZLAres}
\sigma^{(0)}_{\mathrm{is}}-\sigma_D=\frac{e^2}h\frac{x_a^2+x_b^2}{4\mu\tau/\hbar}+\frac{e^2}h\times O[(\mu\tau/\hbar)^{-3}],
\end{equation}
which is confirmed by the computer algebra calculation\cite{theProgram} for the limiting case when \mbox{$2x_ax_b\ll x_a^2+x_b^2\ll1$.}

The absence of CD-loops in the ZLA-diagrams in Fig.~\ref{fZLA} means that the ZLA-contribution neglects interference between electrons.
Thus, the result~\eqref{ZLAres} is valid also for the phase-incoherent system (e.g., at high temperatures).
The ZLA is local -- that is, independent of the macroscopic (on scales $\gg l$) geometrical details of the sample, being the same  in 2D and quasi-1D cases.
Since  the ZLA-diagrams contain no crossings of disorder-averaging lines, their contribution coincides with the results of the
Boltzmann equation approach, see the discussion in~\S9.6 from Ref.~\onlinecite{LevitovShytov}.

Note that \emph{at finite frequency} $\omega$ there are non-zero anisotropic corrections to the conductivity tensor.
We do not present them in the main text; see\cite{theProgram} for details.

According to the loop expansion (see Sec.~\ref{sec:le}), diagrams having one (weak localization) and two loops may produce the contribution to the conductivity
of the same order, or even larger, than~\eqref{ZLAres}. 
We calculate contributions coming from these diagrams in the following sections.

\section{The weak localization contribution\label{sec:WL}}
The weak localization contribution is provided by the diagram in Fig.~\ref{fCondDiags:b} with the renormalized (dashed) Hikami box
\begin{equation}\label{wlde}
\sigma^{(1)}=
\begin{minipage}{8ex}\includegraphics[height=8ex]{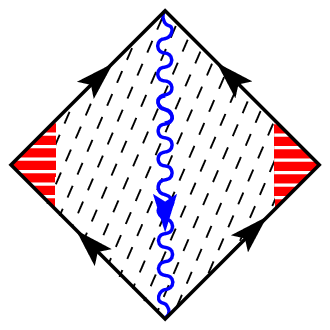}\end{minipage}\equiv
\begin{minipage}{8ex}\includegraphics[height=8ex]{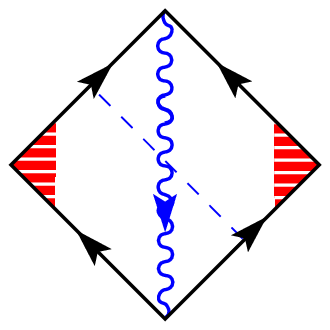}\end{minipage}+
\begin{minipage}{8ex}\includegraphics[height=8ex]{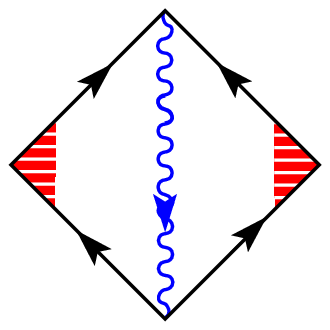}\end{minipage}+
\begin{minipage}{8ex}\includegraphics[height=8ex]{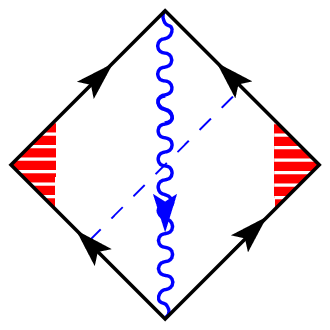}\end{minipage},
\end{equation}
where both vertices are renormalized according to~\eqref{renVelGraph} and~\eqref{renVel}.
The contribution of the diagrams in~\eqref{wlde} can be written in the form
\begin{equation}
\begin{split}
  \sigma^{(1)}_{\alpha\beta}=&\ikd\sum_{i,j=0}^3\mathbb H_{ij}^{\alpha\beta}(\vec k)D^{ij}_{\vec k},\\
\mathbb H_{ij}^{\alpha\beta}&=\mathbb A_{ij}^{\alpha\beta}+\mathbb B_{ij}^{\alpha\beta}+\mathbb C_{ij}^{\alpha\beta},
\end{split}\label{wldeExpr}
\end{equation}
where we used~\eqref{CeqD} and
\begin{widetext}
\begin{equation}\label{wlhb1}
\begin{split}\text{\tt2-1.max}\to\qquad
\mathbb A_{ij}^{\alpha\beta}(\vec k)=
\frac\hbar{2m\tau}\sum_{l=0}^3
\int\frac{\ud^2p_1}{(2\pi\hbar)^2}\Sp_{\mathrm{spin}}
\left[\sigma_lG_R^T(-\vec p_1)\left(-\frac{p_{1\alpha}}m\right)G_A^T(-\vec p_1){\tilde\sigma}_j\GR(\vec p+\vec k)\right]\\\times
\int\frac{\ud^2p_2}{(2\pi\hbar)^2}\Sp_{\mathrm{spin}}
\left[\sigma_lG_R(\vec k-\vec p_2)\frac{k_{2\beta}-p_{2\beta}}mG_A(\vec k-\vec p_2){\tilde\sigma}_i^\dag G_R^T(\vec p_2)\right],
\end{split}\end{equation}
\begin{equation}\label{wlhb2}\text{\tt2-2.max}\to\qquad
\mathbb B_{ij}^{\alpha\beta}(\vec k)=\ipd\Sp_{\mathrm{spin}}
\left[\frac{p_\alpha}m\GR(\vec p){\tilde\sigma}_i^\dag G_A^T(\vec k-\vec p)\frac{k_\beta-p_\beta}mG_R^T(\vec k-\vec p){\tilde\sigma}_j^T\GA(\vec p)\right],
\end{equation}
\begin{equation}\label{wlhb3}\begin{split}
\text{\tt2-3.max}\to\qquad
\mathbb C_{ij}^{\alpha\beta}(\vec k)=
\frac\hbar{2m\tau}\sum_{l=0}^3\int\frac{\ud^2p_1}{(2\pi\hbar)^2}\Sp_{\mathrm{spin}}
\left[\sigma_l^TG_A^T(\vec k-\vec p_1)\frac{k_\beta-p_{1\beta}}mG_R^T(\vec k-\vec p_1){\tilde\sigma}_j^T\GA(\vec p)\right]\\\times
\int\frac{\ud^2p_2}{(2\pi\hbar)^2}\Sp_{\mathrm{spin}}
\left[\sigma_l^TG_A(-\vec p_2)\left(-\frac{p_{2\alpha}}m\right)G_R(-\vec p_2){\tilde\sigma}_i^\dag G_A^T(\vec p+\vec k)\right].
\end{split}\end{equation}
The expressions~\eqref{wlhb1}, \eqref{wlhb2}, \eqref{wlhb3} are generated by our program\cite{theProgram}
and taken from the (automatically created) files {\tt2-1.max}, {\tt2-2.max}, and {\tt2-3.max}.
\end{widetext}

In the absence of orbital dephasing and at zero frequency, the isotropic part of \eqref{wlde} diverges
reproducing the well-known result for the weak antilocalization correction\cite{HikLarkin80,Edelstein:95a,Skvortsov}.

The anisotropic part of \eqref{wlde},\eqref{wldeExpr} converges. Its leading (in the SOI) contribution is given by
\begin{equation}\label{anisWL}
\sigma^{(1)}_{\mathrm{an}}=2x_ax_bS_{20}^0\frac{e^2}h+\frac{e^2}h\times O[(\mu\tau/\hbar)^{-2}],
\end{equation}
where we assumed that $2x_ax_b\ll x_a^2+x_b^2\ll1$, and
\begin{equation}\label{Aexpr}
S_{20}^0=\frac{(131\pi+262\arccot\sqrt7)/\sqrt7-88-7\log2}{224\pi}\approx0.14
\end{equation}
These results are obtained in \cite{theProgram}, where the expression for the renormalized Hikami box (together with other details of calculation) can be found.

The results~\eqref{ZLAres}, \eqref{anisWL} manifest the general rule \eqref{defSan}: the disorder-averaged \emph{conductivity tensor is diagonal} in the considered (rotated by $\pi/4$) basis,
where the SOI is given by~\eqref{neuSOI}. To demonstrate this rule, consider an arbitrary diagram produced by averaging the Kubo formula~\eqref{KuboFormula} for the off-diagonal
conductivity element~$\sigma_{xy}$. Let us change the sign of $p_x$ and of~$\sigma_2$ everywhere in the expression for the diagram. The identity~\eqref{borinoTozhdestvo} remains valid
if the sign of any Pauli matrix is changed, so that expressions for diffusons and cooperons will not change, as well as Hikami boxes, except for the Hikami box with the vertex $v_x$,
which will change sign. Thus the total expression will change its sign; on the other hand, since our transformation is only the change of variables (over which the $\Sp$ is taken), the
expression must remain invariant. So we conclude that \emph{every} diagram for~$\sigma_{xy}$ is zero, and the disorder-averaged conductivity tensor is diagonal.% in the rotated CS.

\section{Two-loop contribution\label{sec:twoLoop}}
\subsection{Expansion in SOI}
In the remainder of the paper we assume that the SOI parameters~\eqref{dRDaAlt} are small, $x\ll1$, $|\delta|\ll1$ (we also assumed this in Sec.~\ref{sec:WL}),
and expand contributions to the conductivity in powers of $x$ and $\delta$. According to Ref.~\onlinecite{anisotrCond}, in an infinite 2DEG,
\begin{equation}
  \begin{split}
    \sigma_{yy}(-a,b)=&\sigma_{xx}(a,b)=\sigma_{xx}(-a,-b)\\
    \text{or}\quad&    \sigma_{yy}(x,-\delta)=\sigma_{xx}(x,\delta).
  \end{split}
\label{deltaSigmaxxyy}
\end{equation}
Then the anisotropic part of the conductivity tensor is given by the following expansion:
\begin{equation}\label{pervoryadDSi}
\sigma_{\mathrm{an}}=\frac{e^2}h\sum_{m,n,r\ge0}S_{mn}^r\frac{x^m\delta^{2n+1}}{(\pf l/\hbar)^r}.
\end{equation}
Physical quantities should depend only on \emph{even} powers of SOI amplitudes; thus, we expect that
$\forall n,r$ $S_{mn}^r=0$ for arbitrary \emph{odd} $m$; our calculations confirm this statement for several values of $n$ and $r$.  The two
loop diagrams can only affect terms with $r\ge1$ in~\eqref{pervoryadDSi}; the calculation shows that $S_{00}^1\ne0$. Below we calculate $\sigma_{\mathrm{an}}$ (i) for a 2D case (an
infinite film -- see Sec.~\ref{sec:2D}), (ii) in quasi-1D case (infinite wire -- see Sec.~\ref{sec:quasi1D}), and (iii) for the quasi-1D ring pierced by
magnetic flux (see Sec.~\ref{sec:brokenTR}).

From Eqs.~\eqref{deltaSigmaxxyy}, \eqref{pervoryadDSi}, and~\eqref{explAN} we conclude that the anisotropic contribution $\sigma_{\mathrm{an}}$ to the conductivity tensor [see the
definition in~\eqref{defSanInv}] can be extracted from the odd in $\delta$ part of $\sigma_{xx}$:
\begin{equation}
  \sigma_{\mathrm{an}}(x,\delta)=\frac{\sigma_{xx}(x,\delta)-\sigma_{xx}(x,-\delta)}2.
\end{equation}

\subsection{The 2D case\label{sec:2D}}
In total, there are nine two-loop diagrams: three are shown in Fig.~\ref{fdivtoPor}, and the other six in Fig.~\ref{fOtherSL}.
The calculation in\cite{theProgram} shows that the leading contribution $S_{00}^1$ is produced by three diagrams in Fig.~\ref{fdivtoPor}.
\begin{figure}
\subfigure[]{\label{fdivtoPor:a}\begin{minipage}{.1\textwidth}\resizebox{\textwidth}{!}{\rotatebox{90}{\input{cc_SOI-b.ins.tex}}}\end{minipage}}\quad
\subfigure[]{\label{fdivtoPor:b}\begin{minipage}{.1\textwidth}\resizebox{\textwidth}{!}{\rotatebox{90}{\input{cc_SOI-c.ins.tex}}}\end{minipage}}\quad
\subfigure[]{\label{fdivtoPor:c}\begin{minipage}{.1\textwidth}\resizebox{\textwidth}{!}{\rotatebox{90}{\input{cc_SOI-d.ins.tex}}}\end{minipage}}
\mycaption{Three relevant two-loop diagrams, which contribute to $S_{00}^1$. See~Ref.\onlinecite{anisotrCond} for more details.
Each diagram contains small-momentum singularities, which mutually cancel each other in accordance with the theorem from Appendix~\ref{app:VW}.
\label{fdivtoPor}}\end{figure}
\begin{figure}
\includegraphics[width=\columnwidth]{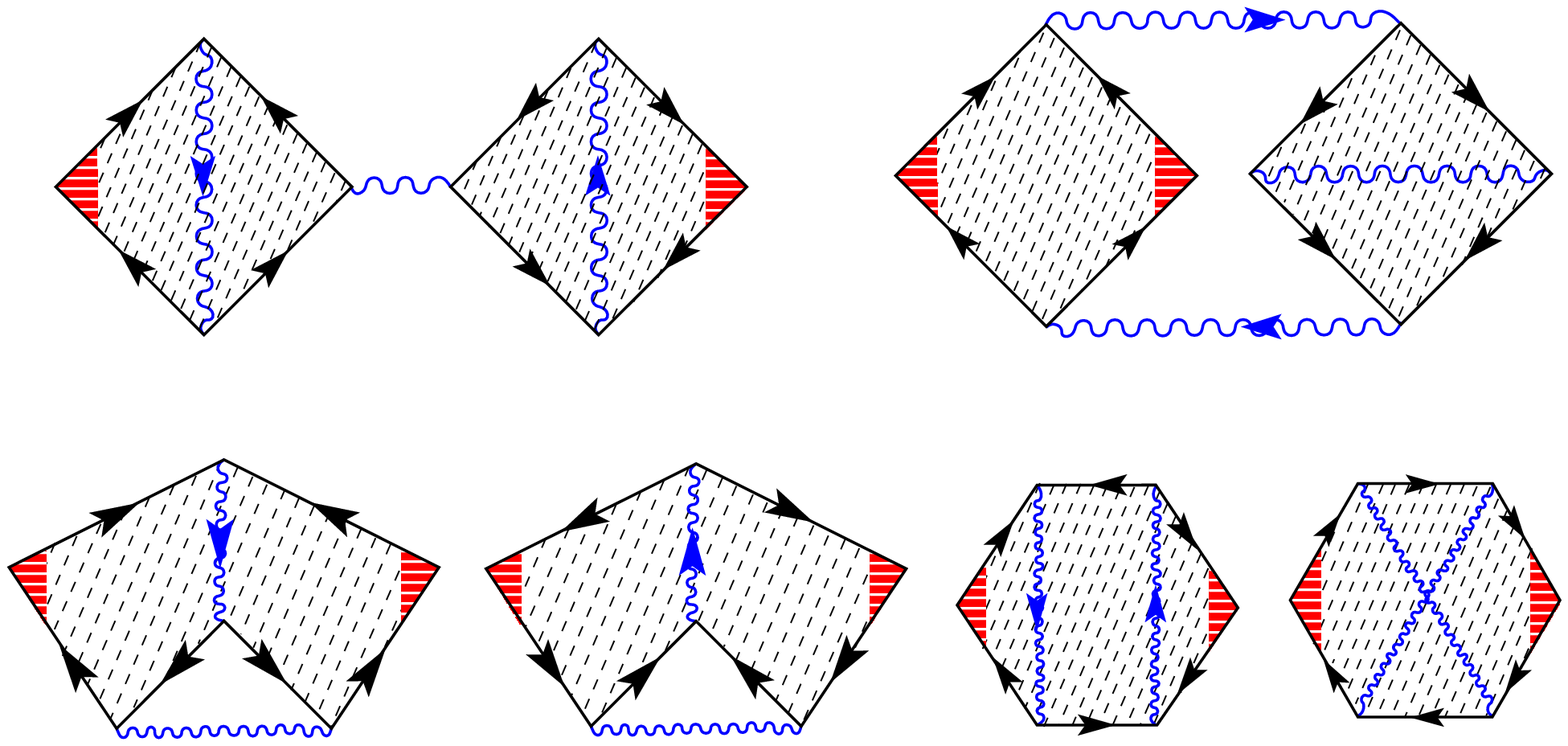}
\mycaption{Six irrelevant two loop diagrams, which do not contribute to $S_{00}^1$.\label{fOtherSL}}\end{figure}
Each relevant diagram consists of two Hikami boxes (dashed) and three diffuson and/or cooperon lines.
For Ref.~\onlinecite{anisotrCond}, we calculated the Hikami boxes manually (that is, we programmed expressions for them \emph{ourselves}
and then computer evaluated them). In order to facilitate the calculation, we made a variable change, which allowed us to express the sum of three diagrams in
Fig.~\ref{fdivtoPor} as the diagram in Fig.~\ref{fdivtoPor:a} with the renormalized upper Hikami box (HB).
We do not use this  trick in this paper because (i) it works only for systems with time-reversal invariance and (ii)
such tricks became useless after we modified the program\cite{theProgram}, which now generates programs for calculating Hikami boxes of arbitrary diagrams.

\begin{figure}
  \centering
  \includegraphics[width=\columnwidth]{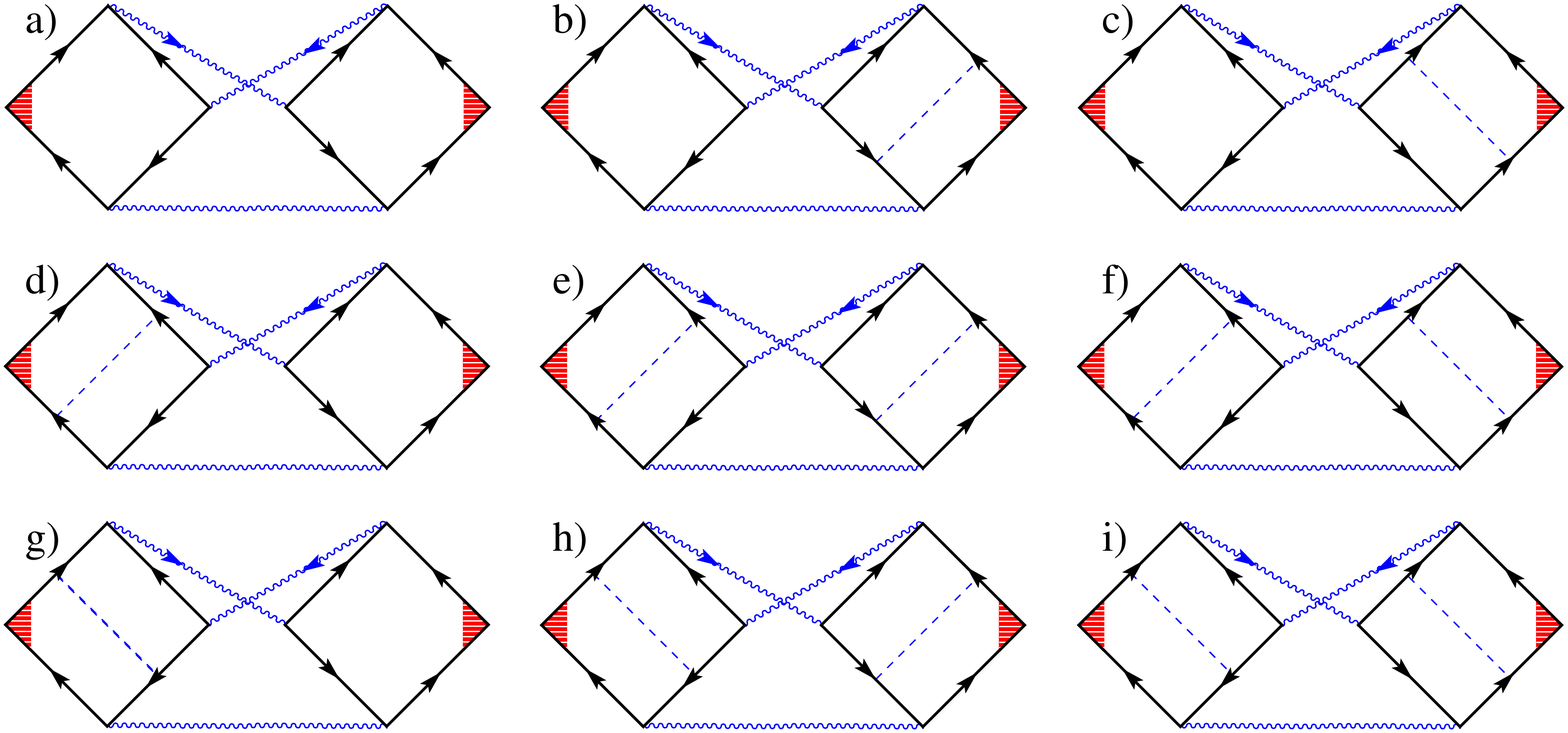}
  \caption{The sum of these nine diagrams with non-dashed Hikami boxes is equal to the diagram in Fig.~\ref{fdivtoPor:c}.}
  \label{fNine}
\end{figure}

Let us now describe how Hikami boxes (HBs) for diagrams in Fig.~\ref{fdivtoPor} are calculated.
Every diagram  in Fig.~\ref{fdivtoPor} contains two dashed HBs, and every dashed HB is given by the sum of three non-dashed HBs, like in Eq.~\eqref{wlde}.
Thus, every diagram  in Fig.~\ref{fdivtoPor} is a sum of \emph{nine} diagrams with non-dashed HBs.
For example, the diagram~\ref{fdivtoPor:c} can be expanded into a sum of nine diagrams in Fig.~\ref{fNine}.
Let us take the diagram~\ref{fNine}d) as an example. It is equal to
\begin{equation}\label{ndeq}  \begin{split}
    \begin{minipage}{.8\columnwidth}\resizebox{\textwidth}{!}{\input{nd.ins.tex}}\end{minipage}\qquad\\
\hspace{-3ex}=\int\frac{\ud^2k_1}{(2\pi\hbar)^2}\int\frac{\ud^2k_2}{(2\pi\hbar)^2}
\mathbb H^{s_1s_2s_3}_{f_1f_2f_3}(\vec k_1,\vec k_2)D^{s_1f_1}_{{\vec k}_1}D^{s_2f_2}_{\vec k_2}D^{s_3f_3}_{\vec k_2-\vec k_1}\\
=\ikd \iqd \mathbb H^{s_1s_2s_3}_{f_1f_2f_3}(\vec k+\vec q,\vec k)D^{s_1f_1}_{\vec k+\vec q}D^{s_2f_2}_{\vec q}D^{s_3f_3}_{-\vec k},
  \end{split}
\end{equation}
where we used~\eqref{CeqD}. The quantity $\mathbb H^{s_1s_2s_3}_{f_1f_2f_3}$ is the product of two HBs; it is  calculated in the (automatically generated) file\cite{theProgram} {\tt batch.me/5-1.max}:
\begin{equation} %\tag{5-1}
 \label{5-1}
  \begin{split}
\mathbb H^{s_1s_2s_3}_{f_1f_2f_3}(\vec k_1,\vec k_2)&=\sum_{i,j=0}^3\frac{\delta_{i,j}}{2m\tau}
    \mathbb A_{is_3f_2}^{\vec k_1\vec k_2}\mathbb B_{js_1}^{\vec k_1\vec k_2}\mathbb C_{f_3f_1s_2}^{\vec k_1\vec k_2},\\
\mathbb A_{is_3f_2}^{\vec k_1\vec k_2}=\Sp_{\vec p}\Sp_{\mathrm{spin}}&\left[\sigma_i\GAT(-\vec p)\sigma_{s_3}^T\right.\\ &\left. \times\GRT(-\vec p+\vec k_2-\vec k_1){\bar\sigma}_{f_2}^T\GA(\vec p+\vec k_1)\right],\\
\mathbb B_{js_1}^{\vec k_1\vec k_2}=\Sp_{\vec p}\Sp_{\mathrm{spin}}&\left[\sigma_j\GA(-\vec p)\left(-\frac{p_x}m\right)\right.\\ &\left. \times\GR(-\vec p){\bar\sigma}_{s_1}^\dag\GAT(\vec p+\vec k_1)\right],\\
\mathbb C_{f_3f_1s_2}^{\vec k_1\vec k_2}=\Sp_{\vec p}\Sp_{\mathrm{spin}}&\left[\frac{p_x}m\GRT(\vec p)\sigma_{f_3}^T\GAT(\vec p-\vec k_2+\vec k_1)\right.\\ &\left. \times{\bar\sigma}_{f_1}\GR(-\vec p+\vec k_2){\bar\sigma}_{s_2}^\dag\GAT(\vec p)\right].
  \end{split}
\end{equation}
These expressions could be simplified, but we on purpose
 wrote them in the same form, as they have been generated in the file {\tt batch.me/5-1.max}
in order to make the comparison easier for an interested reader.

The sum of the three diagrams with dashed HBs in Fig.~\ref{fdivtoPor} is equal to the sum of 27 diagrams with non-dashed HBs.
These HBs are calculated in 27 automatically generated files {\tt batch.me/5-1.max}\ldots {\tt batch.me/5-27.max}

After the HB-calculation, the integration over the diffuson momenta [$\vec k$ and $\vec q$ in~(\ref{ndeq})] is performed.  While the calculation of Hikami boxes
is done fully automatically, integration over diffuson/cooperon momenta must be manually programmed, differently for different problems.  The technical details
are discussed in Appendix~\ref{app:AnalNum}.  The resulting anisotropic contribution to the conductivity is given by~\eqref{pervoryadDSi} for $r=2$ and $m=n=0$:
\begin{equation}
\sigma^{(2)}_{\mathrm{an}}=S^1_{00}\delta\frac{e^2}{2\pi}\frac1{\pf l},\label{sigmaXX}
\end{equation}
where $\delta$ is defined in~\eqref{dRDaAlt} and the coefficient~$S^1_{00}$ is given~by
\begin{equation}
  \mathrm{in\ 2D}\quad S^1_{00}=-5.6\times10^{-3}.\label{finRe}
\end{equation}
The expression~\eqref{sigmaXX} has a non-obvious property
connected with the symmetry of the energy spectrum~\eqref{spektr} with respect to the substitution $a\leftrightarrow b$: the two limiting cases $x_a\ll x_b$ and $x_a\gg x_b$
are described by the same expression~\eqref{sigmaXX}.
Using Eqs.~\eqref{dRDaAlt} and~\eqref{deltaSigmaxxyy}, we rewrite~\eqref{sigmaXX} in the form
\begin{equation}
\sigma^{(2)}_{\mathrm{an}}=S^1_{00}\frac{2x_ax_b}{x_a^2+x_b^2}\frac{e^2}{2\pi}\frac1{\pf l},\quad 2x_ax_b\ll x_a^2+x_b^2\ll1.\label{leadingAnis}
\end{equation}
The contribution~\eqref{leadingAnis} is non-analytic for small $x$: an infinitesimal SOI coupling results in a finite correction $\sigma^{(2)}$ to the
conductivity tensor. We note that a similar non-analyticity occurs also in the weak-antilocalization problem:
if one neglects the dephasing (assuming infinite dephasing length $L_\phi$), an infinitesimal SOI reverts the sign of the
weak-localization correction, switching it to the antilocalization regime.
Similarly to the weak-antilocalization problem, the non-analyticity in~\eqref{finRe} can be smeared by introducing the finite dephasing rate $\tau_\phi^{-1}>0$.
A convenient way to do this is to consider the response for the finite-frequency electric field; see Sec.~\ref{sec:ff}.

\subsection{The quasi-1D case\label{sec:quasi1D}}
Consider a long wire whose cross section $L_\perp$ is much larger than the mean scattering length $l$, but with the corresponding Thouless energy
$E_{c\perp}=\hbar l\vf/2L_\perp^2$ being large compared with the SOI-induced spectrum splitting~\eqref{spectrumSplitting}:
\begin{equation}\label{quasi1D}
E_{c\perp}\tau/\hbar\gg{\tilde x}^2\equiv\max\left(x^2,l^2/L_\phi^2\right).
\end{equation}
When~\eqref{quasi1D} holds, Hikami boxes remain two-dimensional, while diffusons and cooperons become one-dimensional (cf.~\S7.4.1 from
Ref.~\onlinecite{montambauxBook}).  In other words, in~\eqref{5-1} still $\Sp_{\vec p}\equiv\ipds$, but in~\eqref{ndeq}
the momentum integration becomes one-dimensional,
\begin{equation} \label{intQuasi1D}
\ikd\iqd\to\frac1{L_\perp^2}\int_{-\infty}^\infty\frac{\ud k_x}{(2\pi\hbar)^2}\int_{-\infty}^\infty\frac{\ud q_x}{(2\pi\hbar)^2},
\end{equation}
and one should set $k_y=q_y=0$ in the integrand.
From~\eqref{intQuasi1D} one can estimate that the quasi-1D SOI-induced contribution to the conductivity is $l^2/(\tilde xL_\perp)^2\gg1$ times larger than the 2D one.

A quasi-1D sample is macroscopically anisotropic (i.e., does not possess rotational symmetry), but this anisotropy becomes relevant only on the scales much larger than~$l$.
Thus, the Eq.~\eqref{deltaSigmaxxyy}, as well as the claim in Ref.~\onlinecite{anisotrCond} that ``the conductivity tensor is isotropic when $\delta=0$'' are
valid in the ZLA and for the WL-diagram (since the distance between all vertices in corresponding diagrams is of the order of $l$), but may not be valid for the
contribution of the two-loop diagrams in Fig.~\ref{fdivtoPor}.
In fact, we obtain the following anisotropic contribution for the conductivity tensor in the quasi-1D case (in the rotated CS):
\begin{equation}\label{ani1D}
\sigma^{(2)}=\frac{e^2}h\frac\hbar{\pf l}\frac{l^2}{{\tilde x}^2L_\perp^2}
\left[  \begin{pmatrix} -0.39  & 0 \cr0 & 6.7 \end{pmatrix} +  \delta\begin{pmatrix}  -852 & 0 \cr0 & 13 \end{pmatrix}\right],
\end{equation}
which leads to
\begin{equation}\label{qor}
\sigma^{(2)}_{\mathrm{an}}=\frac{e^2}h\frac\hbar{\pf l}\frac{l^2}{{\tilde x}^2L_\perp^2}\left(3.5+433\delta\right).
\end{equation}
Thus, in the quasi-1D case the singularity in the SOI-correction to the conductivity tensor is more pronounced:
(i) it occurs even when the (averaged) energy spectrum is isotropic and (ii)
it diverges at vanishing SOI when orbital dephasing effects are neglected.

\section{The finite frequency case\label{sec:ff}}
In this Section we will see that the anisotropic part of the conductivity tensor
becomes an analytic function of the SOI amplitudes $(x_a,x_b)$ at finite frequency $\omega\ne0$; we assume that the frequency
is large enough compared to the strength of the SOI so that $|\omega\tau|\gg x^2$.

Here we have to expand corrections to the conductivity tensor not only in SOI, but in powers of $|\omega\tau|$ as well.
Similarly to the zero-frequency case we choose three expansion parameters, $x_c$, $x_1$, and $W$, defined in the following way:
\begin{equation} \label{xcxeW} \begin{split}
    x_c=&\sqrt{-2i\omega\tau},\quad x_1=\sqrt{x^2+x_c^2}\ll1,\\ W&=2x_cx/x_1^2,\quad |x_c|,|x_1|,|W|\ll1.
\end{split}\end{equation}

Differently from the zero-frequency case (see Sec.~\ref{sec:CD}), the effect of the SOI on the diffuson can be calculated perturbatively in case when $|\omega\tau|\gg x^2$.
The leading contribution is equal to the diffuson in the absence of SOI:
\begin{equation}\label{dipkc}
D^{\alpha\beta}_{\vec q}=\frac\hbar{m\tau}\frac1{l^2q^2/\hbar^2-2i\omega\tau}\delta_{\alpha\beta},\quad\alpha,\beta=0\ldots3.
\end{equation}
The derivation of SOI-corrections to~\eqref{dipkc} is straightforward, but lengthy, so we do not present it in the text of the article;
see the program\cite{theProgram} for more details.

Comparing~\eqref{dipkc} with expressions for the diffuson at $\omega=0$ (see Secs.~\ref{sec:zeroQ} and~\ref{sec:dbnez}) we see that the diffusons for $|\omega\tau|\ll x^2$ and
$|\omega\tau|\gg x^2$ are very different. Consequently, while at $\omega=0$ the vertex renormalization cancels the anomalous part of the velocity operator, this is no longer the
case for $\omega\ne0$ and at large frequencies the effect of the vertex renormalization is negligible.

Despite that, the calculation of the finite frequency case is similar to the one for $\omega=0$;  differences come from the fact that
now we have three expansion parameters~\eqref{xcxeW} instead of two~\eqref{dRDaAlt} for $\omega=0$.
We obtain\cite{theProgram}
\begin{equation}\label{hochFR} \begin{split}
\sigma^{(2)}_{\mathrm{an}}=-2\cdot0.25\cdot\frac{-2i\omega\tau\cdot2x_ax_b}{(x_a^2+x_b^2-2i\omega\tau)^2}\frac{e^2}{2\pi}\frac1{\pf l},\\
2x_ax_b\ll x_a^2+x_b^2\ll\omega\tau\ll1.
\end{split}\end{equation}
This finite-frequency result obtained first in\cite{anisotrCond} can be interpreted in terms of dephasing; substituting $-i\omega\tau\to\tau/\tau_\phi$, we obtain
\begin{equation}\label{anisRT}
\sigma^{(2)}_{\mathrm{an}}=\left\{\aligned
5.6\times10^{-3}\cdot\frac{\tau_--\tau_+}{\tau_-+\tau_+}\frac{e^2}{2\pi}\frac1{\mu\tau},\quad\tau_\pm\ll\tau_\phi,\\
0.13\cdot\left(\frac{\tau_\phi}{\tau_+}-\frac{\tau_\phi}{\tau_-}\right)\frac{e^2}{2\pi}\frac1{\mu\tau},\quad\tau_\phi\ll\tau_\pm,
\endaligned \right.
\end{equation}
where the Dyakonov-Perel' relaxation times are defined as\cite{spinRelax} $2\tau/\tau_\pm=(x_a\mp x_b)^2$.

\section{The quasi-1D ring pierced by magnetic flux\label{sec:brokenTR}}
The simplest way of breaking the time reversal invariance of the Hamiltonian is considering a constant vector potential field~$\vec A$.
It arises in a quasi-onedimensional ring pierced by a magnetic flux.
(The magnetic flux can be described by subtracting $eA/c$ from momentum arguments of GFs.)
In a ring geometry, the SOI-interaction of the type~\eqref{neuSOI} cannot be provided by usual Rashba and Dresselhaus mechanisms.
However, such SOI is not forbidden and thus can occur due to different reasons, e.g., like in InAs nanowires\cite{Samuelson}.

Let us assume that $\vec A$ is directed along the ring's circumference so that $A\equiv A_{\parallel}\equiv A_x$. The presence of non-zero vector potential breaks the
identities~\eqref{timeRevarsalGF} and~\eqref{CeqD} together with the generalized Vollhardt-Wölfle theorem from Appendix~\ref{app:VW} thus making the anisotropy effect more
pronounced because of the small-momentum diffuson-singularities, which now remain uncompensated.

Eq.~\eqref{CeqD} now changes into
\begin{equation}
C^{\alpha\beta}_{\vec q}=D^{\alpha\beta}_{\vec q-2e\vec A/c}.\label{CeqDmod}
\end{equation}
The calculation of the HBs is the same as in Sec.~\ref{sec:2D} and Sec.~\ref{sec:quasi1D}.
Like in Sec.~\ref{sec:quasi1D} we have to perform summation over two diffuson variables, see~\eqref{intQuasi1D}.
For the diagrams which \emph{contain no cooperons}, this summation is performed in the same way, like for the infinite quasi-1D wire, see Eq.~\eqref{intQuasi1D}.
The diagrams with cooperons become different: in every such diagram, $eA_\parallel/c$ is always subtracted from one (out of two) cooperon momentum.
The summation rule becomes then different from~\eqref{intQuasi1D}:
\begin{equation}
\ikd\iqd\to\frac1{L_\perp^2}\int_{-\infty}^\infty\frac{\ud k^\parallel}{(2\pi\hbar)^2}
\frac1{2\pi L_{\parallel}}\sum_{q^\parallel_n=\frac{2\pi\hbar n}{L_{\parallel}}}
\end{equation}
with $L_{\parallel}$ being the circumference of the ring, and $L_\perp$ its cross section.
The summation is performed over all integer $n$ and it cannot be approximated with integration; one can use the Poisson summation formula instead:
\begin{equation}\label{calcCoop}\begin{split}
\frac1{L_\parallel}&\sum_{n\in{\mathbb Z}}f\left(q^\parallel_n-\frac{2e}cA\right)=\sum_{n\in\mathbb Z}\exp\left[2\pi in\frac\Phi{\Phi_0}\frac e{|e|}\right]C_n,\\
C_n&=\int_{-\infty}^\infty\frac{\ud q^\parallel}{2\pi\hbar}e^{iq^\parallel nL_{\parallel}/\hbar}f\left(q^\parallel\right),\ \Phi=\frac{AL_{\parallel}}c,\ \Phi_0=\frac h{2|e|},
\end{split}
\end{equation}
where the leading ($A$-dependent) contribution comes from the terms with $n=\pm1$,
and we will neglect contribution of terms with $|n|>1$. Keeping only divergent terms (i.e., terms, containing massless cooperon/diffuson matrix elements), and
assuming that $xL_\phi\gg l$ we obtain\cite{theProgram} the flux-dependent correction to the conductivity tensor:
  \begin{equation}\label{piri}
\sigma^{(2)}_{\mathrm{an}}=\frac{e^2}h\left[\cos\left(2\pi\frac\phi{\phi_0}\right)-1\right]\frac{lL_\phi}{\tilde xL_\perp^2}
\frac\hbar{\pf l}(\Sigma_0+\Sigma_1\delta),
  \end{equation}
with the coefficients
\begin{widetext}  \begin{equation}\label{ringAmps}
  \begin{split}
 \Sigma_i=a_ie^{-{\tilde L}_\parallel}+\exp\left[-\frac{{\tilde L}_\parallel}2\sqrt{2\sqrt2-1}\right]
 \left\{b_i\cos\left[\frac{{\tilde L}_\parallel}2\sqrt{2\sqrt2+1}\right]+c_i\sin\left[\frac{{\tilde L}_\parallel}2\sqrt{2\sqrt2+1}\right]\right\},\\
    a_0= -\frac{2{{\tilde L}_\parallel}^2+274{{\tilde L}_\parallel}-219}{128},\quad a_1=-2{{\tilde L}_\parallel}^2-{{\tilde L}_\parallel}-1,\\
b_0=-\frac{7124\sqrt7{{\tilde L}_\parallel}+4513\sqrt{2\sqrt2-1}\sqrt7-2965 \sqrt{2\sqrt2+1}}{1792\sqrt7}=-3{\tilde L_\parallel- 2.1},\quad
b_1=-1.9\cdot10^{-4}{{\tilde L}_\parallel}^3-2.3{{\tilde L}_\parallel}^2-6.7{{\tilde L}_\parallel}+4.2,\\
c_0=\frac{28{{\tilde L}_\parallel}-4513\sqrt{2\sqrt2+1}\sqrt7-2965\sqrt{2 \sqrt2-1}}{1792\sqrt7}=6\cdot10^{-3}{\tilde L}_\parallel+4,\quad
c_1=-8.6\cdot10^{-5}{{\tilde L}_\parallel}^3+4.5{{\tilde L}_\parallel}^2-0.46{{\tilde L}_\parallel} -0.87,
\end{split}  \end{equation}
\end{widetext}
where ${\tilde L}_\parallel=xL_\parallel/l$. Like in Sec.~\ref{sec:ff} we assumed that the divergence at small SOI ($x\to0$) is regularized by the orbital dephasing.
In Eqs.~\eqref{ringAmps} we wrote numerical values of $b_1$ and $c_1$ instead of their analytic expressions in order to save space.
The analytic expressions for $b_1$ and $c_1$ can be found in\cite{theProgram}.

The result~\eqref{piri} has the same order of magnitude in the loop expansion [in powers of~$(\pf l/\hbar)^{-1}$], as (i) the infinite-plane result~\eqref{sigmaXX} and~\eqref{finRe}, as
well as the (ii) quasi-1D result~\eqref{qor}. However,  out of all three considered geometries it is the most sensitive one with respect to orbital dephasing.
In fact, in the coherent limit, $L_\phi\to\infty$, it  diverges as $\propto L_\phi$ for  finite SOI-amplitude $x$, and as $\propto L_\phi^2$ in the limit $x\ll l/L_\phi\to0$.

\section{Conclusions}
We presented symbolic program\cite{theProgram} for generating, sorting, and calculating diagrams in the disorder-averaging diagrammatic technique.
This program strongly facilitates analytical calculations, allowing one to calculate subtle effects due to spin-orbit interaction which were virtually inaccessible
before due to the large number of integrals to be calculated.
The possibility to automatize the calculation improves the usefulness of the diagrammatic approach, especially also in comparison to the non-linear $\sigma$-model\cite{Kamenev2}, as a tool for studying disordered systems.

Using this program, we studied anisotropic corrections to the conductivity tensor due to the spin-orbit interaction (SOI).
The arising anisotropy is a phase-coherence effect; therefore it strongly depends on the geometry of the sample.
In the quasi-1D wire the anisotropic correction is larger than in an infinite 2D-plane.
Moreover, while in 2D-case the effect arises due to the anisotropy of the energy spectrum induced by the interference between Rashba and Dresselhaus types of SOI,
in the quasi-1D case the conductivity is anisotropic even in the presence of only one type of SOI (Rashba or Dresselhaus), that is, when the energy spectrum is isotropic.
The (microscopic) anisotropy of the conductivity tensor arises due to the macroscopic (shape) anisotropy of the sample (on the scale much larger than the mean free path~$l$).

We also studied the case when the time-reversal symmetry of the system is broken (a ring pierced by a magnetic flux); then the anisotropy of the conductivity tensor
becomes more sensitive to orbital dephasing effects  due to the uncompensated small-momentum divergences in the integration over the diffuson momentum.

In all the considered geometries, the effect is non-analytical in the amplitude of the spin-orbit interaction, if the orbital dephasing effects are not taken into
account. Once the dephasing effects are considered, the conductivity becomes an analytical function of the spin-orbit amplitudes.

We are grateful to M.~Duckheim and D.~Maslov for helpful discussions.
We acknowledge financial support from the Swiss NSF and the NCCR Nanoscience.

\appendix

\begin{widetext}
\section{Inhomogeneous distribution of electric field in a homogeneous wire\label{app:leadingCond}}
The Kubo formula~\eqref{KuboFormula} is not valid for the calculation of the leading (Drude) contribution to the conductivity.
From Sec.~\ref{sec:Kubo} one obtains:
\begin{equation}\label{ZLAdivergence}
\overline{\Sp\left[\delta{\hat\rho}_N{{\hat j}^{\,\alpha}}_N\right]}=
-\frac{ie^2A_\omega^\alpha\omega}{c\hbar m^2}\iE f_E'\overline{\Sp\left[{\hat p}_\alpha^2\hgre\hgae\right]},\quad\omega\to0,
\end{equation}
where the main contribution to $\overline{\Sp\left[{\hat p}_\alpha^2\hgre\hgae\right]}$ is given by $\Sp\left[{\hat p}_\alpha^2\gre\gae\right]$.
(Note that we do not take into account spin degree of freedom in this section and use expressions~\eqref{agf} for Green functions; consequently, $\Sp$ operators here do not contain
  trace over spin.)

Only values of $E$ close to the Fermi level ($E=0$) contribute to the integral in~\eqref{ZLAdivergence}, so that
\begin{equation}\label{}
\Sp\left[{\hat p}_\alpha^2\gre\gae\right]=2im\tau\,\Sp\left[\frac{{\hat p}_\alpha^2}{2m}\left(\gre-\gae\right)\right]
=2im\tau\,\overline{\Sp\left[\frac{{\hat p}_\alpha^2}{2m}\left(\hgre-\hgae\right)\right]}.
\end{equation}
From the Lehmann representation we obtain the matrix element of $\hgre-\hgae$ in some (arbitrary) $\lambda$-representation:
\begin{equation}\label{densityGRA}
\langle\lambda|\hgre-\hgae|\lambda'\rangle =-2\pi i\sum_n\delta(E-E_n)\langle\lambda|n\rangle\langle n|\lambda'\rangle,\quad E_n=\langle n|\hat H|n\rangle,
\end{equation}
where $|n\rangle$ are the exact eigenstates of the unaveraged Hamiltonian.
In a (spinless) disordered system, the states $|n\rangle$ are non-degenerate due to the disorder;
thus one can rewrite~\eqref{densityGRA} in the $|\epsilon\rangle$-basis with $|n\rangle\equiv|\epsilon\rangle$:
\begin{equation}\label{rss}
\Sp\left[\frac{{\hat p}_\alpha^2}{2m}\left(\hgre-\hgae\right)\right]
=\frac1d\sum_\epsilon\langle\epsilon|\frac{{\hat p}^2}{2m}|\epsilon\rangle
\langle\epsilon|\hgre-\hgae|\epsilon\rangle
=-\frac{2\pi i}d\sum_{\epsilon,\epsilon'}\delta(E-\epsilon')
\langle\epsilon|\frac{{\hat p}^2}{2m}|\epsilon\rangle
\langle\epsilon|\epsilon'\rangle\langle\epsilon'|\epsilon\rangle
=-\frac{2\pi i}d\sum_{\epsilon'}\delta(E-\epsilon')\langle\epsilon'|\frac{{\hat p}^2}{2m}|\epsilon'\rangle,
\end{equation}
where $d$ is the dimension ($d=2$ in 2D). Substituting~\eqref{rss} into~\eqref{ZLAdivergence} we obtain
\begin{equation}
j^\alpha(\omega)=\sigma(\omega)\frac{i\omega}cA^\alpha_\omega=
\overline{\Sp\left[\delta{\hat\rho}_N{{\hat j}^{\,\alpha}}_N\right]}
=-\frac{2ie^2A\omega\tau}{mcd\hbar}\int\ud E\nu f_E'\langle E|\frac{{\hat p}^2}{2m}|E\rangle,\quad\omega\to0.
\end{equation}
Thus the main (Drude) contribution to the conductivity \emph{per spin projection} is given by
\begin{equation}\label{inhCond}
\sigma_D=-\frac{2e^2}{dm\hbar}\tau\int\ud E\nu f_E'\langle E|\frac{{\hat p}^2}{2m}|E\rangle.
\end{equation}
The easiest case is equilibrium at $T=0$: then $f_E'=-\delta(E)$ and the integral is equal to $\nu\langle0|\frac{{\hat p}^2}{2m}|0\rangle\equiv\nu\mu$,
so that $\sigma_D=\mu\tau\cdot e^2/h\cdot4\pi\nu/dm$.

Consider a wire of length $L$ between two leads under the voltage $V$, so that the energy distribution in the leads is given by $f_E^{R,L}$.
If the effect of the interaction between electrons in the wire is weak, its distribution function linearly depends on the coordinate:\cite{Kamenev2}
\begin{equation}
f_E(r)=\frac rLf_E^L+\left(1-\frac rL\right)f_E^R.
\end{equation}
From~\eqref{inhCond} we conclude that also the leading (Drude) contribution to the conductivity is slightly inhomogeneous.
Consequently, the stationary distribution of the electric field in the wire will also be inhomogeneous, so that the wire will be homogeneously charged.
Thus the charge-neutrality of the current-currying wire is slightly violated.
\end{widetext}

\section{Generating diagrams on computer\label{app:program}}
The diagrams are generated in the program\cite{theProgram} from the Kubo formula~\eqref{KuboFormula} according to the following algorithm.
\begin{enumerate}
\item The simplest diagram is obtained from~\eqref{KuboFormula} by substituting GF-operators (${\hat G}_{\mathrm R}^E$ and ${\hat G}_{\mathrm A}^{E-\omega}$)
  with their averaged values. The result corresponds to the diagram in Fig.~\ref{fZLA:a}.
\item Add one cooperon or one diffuson line to the diagram(s) obtained on the previous step in all possible ways.
\item Leave only diagrams having no more than two loops (to calculate the number of loops, the diagram has to be redrawn in the ``coordinate representation'',
  see Sec.~\ref{sec:twoWays}).
\item Recursively perform two previous steps, until the last step produces no more new diagrams.
\item Redraw all diagrams in the ``coordinate representation''. Add single impurity lines in the Hikami boxes (in all possible ways), cf.~\eqref{wlde}.
\item E.g., consider the diagram in Fig.~\ref{fdivtoPor:a} without lonely disorder lines inserted in the Hikami boxes;
 each of its three diffuson lines is given by the infinite series~\eqref{ids}. Leaving only the
  first terms in these series, we note that the resulting diagram is exactly the one depicted in Fig.~\ref{fCondDiags:a} with three disorder (dashed) lines
  left. We see that the diagram in Fig.~\ref{fdivtoPor:a} includes the contribution which has been already taken into account in the weak localization
  diagram. To prevent double-counting, it should be subtracted. Such situation (when two different diagrams have common contributions) can occur also for many
  other diagrams. All the diagrams must be checked for that; double-counted contributions must be subtracted.
\end{enumerate}
In principle, the above steps can be done manually, but it is better to make use  of the computer program\cite{theProgram}.
On the first step we started with one diagram; in the end we get 215 ones.\cite{note:dashedHB}

\section{$S_{00}^1$ at $\omega=0$: calculating $\ikds\iqds$ on computer: analytics and numerics\label{app:AnalNum}}
\setcounter{equation}{0}\renewcommand{\theequation}{\Alph{section}\arabic{equation}}
Once all diagrams with no more than two CD-loops are generated, they are automatically divided into different groups according to the number of HBs and diffuson
lines (in total we get eight groups). The Hikami boxes of all diagrams are automatically calculated.
We have checked that the diagrams of only one (the fifth) group contribute to $\sigma_{\mathrm{an}}$.
The diagrams of the fifth group are calculated in the directory {\tt 2D.5/.}

Within the same group, the Hikami boxes can be summed up; after that the expression for the sum of all diagrams from the group can be written in the form~\eqref{ndeq}, with
$H^{s_1s_2s_3}_{f_1f_2f_3}$ being the sum of all HBs.

Then we integrate over the diffuson momenta.
We use dimensionless momentum variables ($\vec K$ and $\vec Q$) defined in~\eqref{defBP}.
We checked\cite{theProgram} that only the diagrams in Fig.~\ref{fdivtoPor} contribute to $S_{00}^1$.
Each of these diagrams has three diffuson/cooperon lines; we label their momenta as $\vec K$, $\vec Q$, and $\vec K+\vec Q$.
In case of small anisotropy of the energy spectrum, $\delta\ll1$, the denominator of the diffuson/cooperon depends only on the modulus of its momentum.
Thus the angular dependence of the denominator comes only from $(\vec K+\vec Q)^2=K^2+Q^2+2KQ\cos\psi$, where $\psi$ is the angle between $\vec K$ and $\vec Q$.

We calculate $\ikds\iqds$ in polar coordinates. In total there are two angular integrations; since the denominator depends only on one angle, another angular
integration can be easily (and analytically) performed.
The second angular integration is more complicated: the integrand is given by the sum of rational functions which have the form
\begin{equation}
\int_0^{2\pi}\frac{\ud\psi}{2\pi}\frac{P_1(\sin\psi,\cos\psi)}{P_2(\cos\psi)},
\end{equation}
where $P_{1,2}$ are polynomials. The denominator is even in $\psi$, so we can leave only even (in $\psi$) part of the numerator.
Then numerator thus can be expressed as a another polynomial: % $P_3(\cos\phi)$, so that our integral now looks like
$$
\int_0^{2\pi}\frac{\ud\psi}{2\pi}\frac{P_3(\cos\psi)}{P_2(\cos\psi)},\quad P_3(\cos\psi)=\sum_{n\ge1}a_n\cos^n\psi.
$$
Next, we perform the analytical integration over $\psi$.
Because of the large number of terms to integrate and
large size of the expressions, this analytical integration (basically, calculation of residues) can only be done on computer.
We use the fact that the denominators of all massfull elements of the diffuson $D_{\vec k+\vec q}$ can be factorized into two expressions.
In the program\cite{theProgram} they are denoted as
{\tt uno}$=1+(\vec K+\vec Q)^2$ and {\tt due}$=2-(\vec K+\vec Q)^2+(\vec K+\vec Q)^4$. The size of the integration result grows rapidly with powers of {\tt uno}
and {\tt due} in the denominator, so it is necessary to split integrands into elementary fractions.\cite{note:falseDiverg}

We assume that integrals over $(\vec k,\vec q)$ converge on the scale of $k\lesssim x\hbar/l$ and $q\lesssim x\hbar/l$, so that the integrals in dimensionless variables
$(\vec K,\vec Q)$  converge on the scale of $K\lesssim$ and $Q\lesssim1$.
Together with the assumption $x\ll1$, this permits us to use Taylor expansions (e.g., for Hikami boxes) in powers of $kl/\hbar\ll1$ and $ql/\hbar\ll1$.
The usage of Taylor expansion here corresponds to the diffusive limit and is justified by the fact that we do not get any large-momenta divergences.

Next, the integrand is symmetric with respect to $K\leftrightarrow Q$, and the integration operator has this symmetry too.
So we symmetrize every term of the integrand, and express it in terms of new variables $P=K^2+Q^2$ and $A=2KQ/P$.
Accordingly, our integration operator is changed:
\begin{equation}\label{fvc} \begin{split}
\int_0^\infty\!\!\ud K\int_0^\infty\!\!\ud Q\,KQ=2\int_0^\infty\!\!\ud K\int_0^K\!\!\ud Q\,KQ\\
=2\int_0^\infty\!\!\!\!\ud P\int_0^1\!\!\ud A\frac{PA}{2J},\quad A=\frac{2KQ}{K^2+Q^2},
\end{split}\end{equation}
and $J$ is the Jacobian:
\begin{equation}
J=\left|\frac{\partial(P,A)}{\partial(K,Q)}\right|=\left|4\frac{K^2-Q^2}{K^2+Q^2}\right|=4\sqrt{1-A^2}.
\end{equation}
The integration results produced by {\tt integrate.max} for $A\to0$ have insufficient precision
(since the difference of large numbers has low numerical accuracy), so we integrate by $P$ and $A$ not from zero but from $0.001$.
Since $\forall P$ the integrand$=0$ at $A=0$, and maximal contribution to the result occurs for $A\to1$, this adjustment of the lower limit  introduces only
a negligible error into the result.\cite{note:As}

We treat separately (see files {\tt KpQ.max}) the terms containing massless elements of $D_{\vec k+\vec q}$, and see that no divergences at small $(P,A)$ occur,
as it is predicted according to the VW-theorem in Appendix~\ref{app:VW}.
\begin{figure}
\resizebox{.35\textwidth}{!}{\input{RDsoiInt.ins.tex}}
\mycaption{The integrand for $\int_0^1\frac{\ud A}{\sqrt{1-A}}$.\label{fIntegrandForIdA}}
\end{figure}

The $\int_0^\infty\ud P$ together with the subsequent $\int_0^1\ud A$ is performed in directory {\tt2D.5/A/}.
(As a check, we performed the numerical integration using alternative variables in the directory  {\tt2D.5/B/}, which led to the same result.)
We use functions {\tt qagi} for $\int_0^\infty\ud P$ and {\tt qaws}  for $\int_0^1\ud A$, which are part of the
\inslna{http://www.netlib.org/quadpack/}{{\tt quadpack}} package\cite{quadpack}.
In order to minimize possible rounding corrections, all calculations are done with high precision (35 digits).

The integrand for $\int_0^1\frac{\ud A}{\sqrt{1-A}}$ is plotted in Fig.~\ref{fIntegrandForIdA}.
The final result~\eqref{finRe} has been calculated using different {\tt lisp} realizations ({\tt clisp}, {\tt gcl} and  {\tt sbcl}), and for different
numbers of digits in the numerical integration (20, 25, and 35).

In conclusion, the steps of the diffuson momentum integration are:
\begin{enumerate}
\item Perform the first (easy) angular integration.
\item Select terms that contain massless component $D^{00}_{\vec k+\vec q}$. Integrate them by $\psi$ (second angular integration variable).
\item Split other terms into elementary fractions, in order to decrease powers in their denominators. Integrate them by $\psi$, and combine them with the terms
  obtained in the previous step.
\item Calculate the integrand in many points within the interval $A\in[0,1]$.
Interpolate (linearly) the integrand in the integration interval $A\in[0,1]$, and integrate the interpolation result by~$A$.
\end{enumerate}

\section{Generalization of the Vollhardt-Wölfle theorem for Hamiltonians with spin-orbit interaction\label{app:VW}}
\setcounter{equation}{0}\renewcommand{\theequation}{\Alph{section}\arabic{equation}}
We assume that the following statements (demonstrated above for the case of Rashba and Dresselhaus SOI) are valid in general for a system with time reversal symmetry:
\begin{itemize}
\item The anomalous part of the velocity operator vanishes at $\omega=0$ due to the vertex renormalization~\eqref{renVel}.
\item \emph{The only}\cite{note:withoutSOI} divergent (massless) element of the diffuson matrix~\eqref{defXabDC} is $D^{00}$.
\end{itemize}
Then, the Vollhardt-Wölfle theorem\cite{woelfe} can be generalized to spin-dependent Hamiltonians:
\begin{thma}\label{VW}
 In the the calculation of linear response coefficients,  no diffuson-type divergences occur if the unperturbed system possesses time reversal symmetry.
\end{thma}
In particular, diffuson singularities cannot occur in the calculation of the conductivity or spin susceptibility, as well as in the calculation of corresponding
cumulants.

\begin{figure}
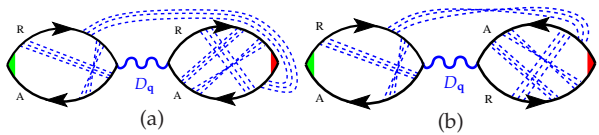

  \centering
\subfigure[]{\label{fNasty}\begin{minipage}{.45\columnwidth}\resizebox{\textwidth}{!}{\input{nasty.ins.tex}}\end{minipage}}
\subfigure[]{\label{fSister}\begin{minipage}{.45\columnwidth}\resizebox{\textwidth}{!}{\input{sister.ins.tex}}\end{minipage}}
  \caption{Illustration to the proof of the Theorem~\ref{VW}.
Only one diffuson (with momentum $\vec q$) is depicted in the ``coordinate representation''; this diffuson $D_{\vec q}$ divides both diagrams in two
parts (bubbles). The vertex on the rhs of both diagrams is proportional to the renormalized velocity operator~\eqref{renVel}. The rhs-bubble of (b) is the mirrored (and then
rotated by 180\ensuremath{^\circ}) rhs-bubble of (a). The sum of two diagrams is regular at $q\to0$.}
  \label{fNastyDiagram}
\end{figure}

We consider an arbitrarily complicated diagram with one or more diffusons.
We choose any of them and prove that the coefficient in front of $D^{00}$ vanishes when the momentum flowing through this diffuson approaches zero.
Let us draw only the selected diffuson in the  ``coordinate representation'' (see Sec.~\ref{sec:twoWays}); other diffusons and all cooperons (if present) remain
in the ``ladder representation''; e.g., Fig.~\ref{fNastyDiagram}.
Drawn in this way, a diagram consists of two bubbles with a wavy diffuson line between them [e.g., Fig.~\ref{fNasty}].
In a diagram for linear response to applied electric field, (at least) one of two vertices (we assume that it is the rhs-one) is proportional to the
renormalized velocity operator~\eqref{renVel}. Let us downgrade all diffusons/cooperons to crosses [removing  disorder-averaging (dashed) lines connecting them].\cite{note:crosses}
After that the rhs-bubble of an arbitrary diagram can be written as
\begin{eqnarray}\label{prapuz}
\lefteqn{\begin{minipage}{20ex}\resizebox{\textwidth}{!}{\input{rhs-bub.ins.tex}}\end{minipage}}\\ & &
   \propto\ipd{\color{red}\vec p\vec A}\Sp_{\mathrm{spin}}
\left[{\color{blue}\sigma_0}G_{\mathrm R}^{(m)}(\vec p-\vec q,\vec p)G_{\mathrm A}^{(n)}(\vec p,\vec p-\vec q)\right],
\nonumber
\end{eqnarray}
where $G_{\mathrm R}^{(m)}(\vec p-\vec q,\vec p)$ and $G_{\mathrm A}^{(n)}(\vec p,\vec p-\vec q)$ are \emph{unaveraged} Green functions in the $m$th and $n$th
order of the perturbation theory in the disorder potential.
[Note that we did not draw the crosses on the GFs-lines in~\eqref{prapuz}.]
Since the disorder cannot break the time reversal symmetry of the unperturbed Hamiltonian,
\begin{equation}
\hspace{-2ex}\forall m,n,\vec p_1,\vec p_2\ \left\{\aligned
    \sigma_2\left[G_{\mathrm R}^{(m)}(-\vec p_2,-\vec p_1)\right]^T\sigma_2=&G_{\mathrm R}^{(m)}(\vec p_1,\vec p_2),\\
    \sigma_2\left[G_{\mathrm A}^{(n)}(-\vec p_2,-\vec p_1)\right]^T\sigma_2=&G_{\mathrm A}^{(n)}(\vec p_1,\vec p_2),
\label{timeRevarsalGFdi}\endaligned \right.
\end{equation}
which is a generalization of~\eqref{timeRevarsalGF}. From our assumptions it follows that
$\GRA$ commutes with the renormalized velocity vertex $\vec v\vec A$. Applying the transformation~\eqref{timeRevarsalGFdi} to GFs in~\eqref{prapuz}, transposing
matrices under the trace, and substituting $\vec p\to -\vec p$ in the integral we obtain
\begin{equation}
\begin{split}
\hspace{-2ex}-\ipd\vec p\vec A\Sp_{\mathrm{spin}}
\left[\sigma_\beta G_{\mathrm R}^{(m)}(\vec p-\vec q,\vec p)G_{\mathrm A}^{(n)}(\vec p,\vec p-\vec q)\right]&\\
\hspace{-2ex} =\ipd\vec p\vec A\Sp_{\mathrm{spin}}
\left[\sigma_\beta G_{\mathrm R}^{(m)}(\vec p,\vec p+\vec q)G_{\mathrm A}^{(n)}(\vec p+\vec q,\vec p)\right],&
  \end{split}
\end{equation}
so that~\eqref{prapuz} vanishes at $q=0$, and the diffuson divergence is regularized. The proof can be performed for every diffuson in the diagram, so that
all diffusons are regularized.
An example of mutually cancelling diffuson-divergences is depicted in Fig.~\ref{fNastyDiagram}: both diagrams \ref{fNasty} and~\ref{fSister} diverge at
$\vec q\to0$, but their sum is regular.
A similar cancellation takes place for the diagrams in Fig.~\ref{fCoopDiff}:
the singularity in $D^{00}_{\vec k}$ in the diagram~\ref{fdivtoPor:a} is cancelled (i.e., regularized) by the diagram~\ref{fdivtoPor:b}, while
the singularity in $D^{00}_{\vec q}$ in the diagram~\ref{fdivtoPor:a} is cancelled by the diagram~\ref{fdivtoPor:c}.
Such cancellation is provided by the fact that \emph{in a system with time-reversal invariance}
cooperon components are \emph{exactly} the same as components of a diffuson. (This remains true also in the presence of interaction between electrons.\cite{castellani})

Thus we demonstrated that (in the absence of interaction) there can be no diffuson type singularities in the presence of the time-reversal symmetry, thus
generalizing the theorem proved in Ref.~\onlinecite{woelfe} for the spinless case without SOI. 

%\bibliography{refs.aps,books.aps,local.aps}

\begin{thebibliography}{43}
\bibitem[{not({\natexlab{a}})}]{note:noDisp}
\bibinfo{note}{Note that the spatial dispersion of the conductivity can be
  neglected, so that the relation between the applied electric field and
  induced current density is local in space.}
\bibitem[{not({\natexlab{b}})}]{note:altPauliMatr}
\bibinfo{note}{In\cite{anisotrCond} the SOI \eqref{neuSOI} and all consequent
  equations are written using an alternative set of Pauli matrices. The
  introduction of such alternative set is unnecessary; in this paper we use
  standard Pauli matrices}.
\bibitem[{not({\natexlab{c}})}]{note:NUC}
\bibinfo{note}{In particular, one approximation that neglects non-universal
  corrections is $\ipds\approx\nu\ixi$. We checked that this aproximation does
  not affect our results.}
\bibitem[{not({\natexlab{d}})}]{note:Drude}
\bibinfo{note}{Just like Eq.~\eqref{KuboFormula}, the diagram in
  Fig.~\ref{fZLA:a} can not be used for the calculation of the leading (Drude)
  contribution to the conductivity $\sigma_D$. However, it can be used for
  calculating correstions to $\sigma_D$, see Sec.~\ref{sec:ZLA}.}
\bibitem[{not({\natexlab{f}})}]{note:dashedHB}
\bibinfo{note}{Note that this number is decreased if one uses the ``dashed
  Hikami box'' notation, like in Fig.~\ref{fdivtoPor}, where every Hikami box
  is given by the sum of three, cf.~\eqref{wlde}. So, every diagram in
  Fig.~\ref{fdivtoPor} represents nine ``elementary'' diagrams, so that three
  diagrams in Fig.~\ref{fdivtoPor} correspond to 27 ``elementary'' diagrams out
  of total 215.}
\bibitem[{not({\natexlab{g}})}]{note:falseDiverg}
\bibinfo{note}{If the splitting of the integration result into elementary
  fractions is not performed, the subsequent numerical integration is
  problematic. This happens because huge integration results usually contain
  the difference of large numbers, which is calculated numerically with low
  accuracy, and result in various numerical errors, like, e.g., ``false''
  divergences.}
\bibitem[{not({\natexlab{h}})}]{note:As}
\bibinfo{note}{The asymptotic for $A\ll1$ can be easily calculated: one expands
  $1/uno$ and $1/due$ in small $A$; then $\int_0^{2\pi}\ud\psi/(2\pi)$ becomes
  trivial, since $\psi$ appears now only in the numerator, and we get a
  convenient integrand for $\int_0^1\ud A\int_0^\infty\ud P$.}
\bibitem[{not({\natexlab{i}})}]{note:withoutSOI}
\bibinfo{note}{This condition is not fulfilled for a spin-independent
  Hamiltonian: $D^{\alpha\beta}\propto D^{00}\delta_{\alpha\beta}$, so that all
  four non-zero matrix elements diverge at zero frequency and momentum.
  However, $(D^{11},D^{22},D^{33})$ come with zero coefficient so that the
  theorem remains valid}.
\bibitem[{not({\natexlab{j}})}]{note:crosses}
\bibinfo{note}{This, e.g., would correspond to elimination of dashed lines and
  leaving only crosses in Fig.~103 in Ref.~\onlinecite{AGD}.}
\end{thebibliography}

\end{document}